\documentclass[conference, a4paper]{IEEEtran}

\usepackage{etoolbox}
\makeatletter
\def\ps@headings{%
	\def\@oddhead{\mbox{}\scriptsize\rightmark \hfil \thepage}%
	\def\@evenhead{\scriptsize\thepage \hfil \leftmark\mbox{}}%
	\def\@oddfoot{}%
	\def\@evenfoot{}}
\makeatother
\usepackage{color}
\usepackage{amsfonts}
\usepackage{mathrsfs}
\usepackage{amsfonts}
\usepackage{amssymb}
\usepackage{graphicx}
\usepackage{epsfig}
\usepackage{epstopdf}
\usepackage{psfrag}
\usepackage{amsmath}
\usepackage{array}
\usepackage{cases}
\usepackage{subfigure}
\usepackage{cite}
\usepackage{algorithmic}
\usepackage{algorithm}

\usepackage{lipsum}
\usepackage{url}
\usepackage{enumerate}

\IEEEoverridecommandlockouts

\newtheorem{theorem}{Theorem}
\newtheorem{lemma}{Lemma}

\newtheorem{Exam}{Example}

\newtheorem{problem}{Problem}

\newcommand{\blue}{\textcolor{black}}

\begin{document}
\bibliographystyle{IEEEtran}
	
\title{Optimal Multi-View Video Transmission in Multiuser Wireless Networks by Exploiting Natural and View Synthesis-Enabled Multicast Opportunities}

\author{Wei Xu, \quad Ying Cui, \quad Zhi Liu\thanks{Manuscript received 
		April~10, 2019; revised August~20, 2019 and  November~2, 2019; accepted November~14, 2019.
		The work of Y. Cui was supported in part by NSFC China (61771309, 61671301, 61420106008, 61521062). The work of Z. Liu was supported by JSPS KAKENHI grants 18K18036, 19H04092, and The Telecommunications Advancement Foundation Research Fund. The paper has been presented in part at the IEEE GLOBECOM 2018\cite{xu2018energy}. The associate editor coordinating the review of this paper and approving it for publication was Prof. Asaf Cohen. (Corresponding author: Ying Cui.) 
		
	    W. Xu and Y. Cui  are with the Department of Electronic Engineering, Shanghai Jiao Tong University, Shanghai 200240, China (e-mail:cuiying@sjtu.edu.cn).
	    
	    Z. Liu is with the Department of Mathematical and Systems Engineering, Shizuoka  University, Hamamatsu 432-8561, Japan (e-mail:liu@ieee.org). }}

\pagestyle{headings}

\maketitle
\begin{abstract}
Multi-view videos (MVVs) provide immersive viewing experience, at the cost of traffic load increase for wireless networks. In this paper, we would like to optimize MVV transmission in a multiuser wireless network by exploiting both natural multicast opportunities and view synthesis-enabled multicast opportunities. Specifically, we first establish a mathematical model to specify view synthesis at the server and each user, and characterize its impact on multicast opportunities. This model is highly nontrivial and fundamentally enables the optimization of view synthesis-based multicast opportunities. For given video quality requirements of all users, we consider the optimization of view selection, transmission time and power allocation to minimize the average weighted sum energy consumption for view transmission and synthesis. In addition, under the energy consumption constraints at the server and each user respectively, we consider the optimization of view selection, transmission time and power allocation and video quality selection to maximize the total utility. These two optimization problems are challenging mixed discrete-continuous optimization problems. For the first problem, we propose an algorithm to obtain an optimal solution with reduced computational complexity by exploiting optimality properties. For each problem, to reduce computational complexity, we also propose a low-complexity algorithm to obtain a suboptimal solution, using Difference of Convex (DC) programming. Finally, numerical results show the advantage of the proposed solutions over existing ones, and demonstrate the importance of the optimization of view synthesis-enabled multicast opportunities in MVV transmission.
\end{abstract}


%
\IEEEpeerreviewmaketitle
\section{Introduction}
A multi-view video (MVV) is generated by capturing a scene of interest with multiple cameras from different angles simultaneously. Each camera can capture both texture maps (i.e., images) and depth maps (i.e., distances from objects in the scene), providing one view. Besides views captured by cameras, additional views, providing new view angles, can be synthesized based on reference views using Depth-Image-Based Rendering (DIBR) \cite{fehn2004depth}. A MVV subscriber (i.e., user) can freely select among multiple view angles, hence enjoying immersive viewing experience. MVV is one key technique in free-viewpoint television, naked-eye 3D and virtual reality (VR) \cite{4037061,1369701}. Thus, it has vast applications in entertainment, education, medicine, etc. The global market of VR related products is predicted to reach 30 billion USD by 2020 \cite{123456}.

A MVV is in general of a much larger size than a traditional single-view video, bringing a heavy burden to wireless networks. Multiple views of a MVV can be jointly encoded using multiview video coding\cite{vetro2011overview,4378926} or separately encoded using state-of-the-art codec such as H.264/AVC and HEVC\cite{wiegand2003overview}. In particular, joint encoding achieves a significant coding gain by exploiting statistical dependencies from both temporal and inter-view reference frames for motion-compensated prediction \cite{fujihashi2014umsm,de2015optimizing,de2013optimized,merkle2007efficient}. 
However, it yields a great traffic load causing bandwidth waste. This is because with joint encoding, multiple views have to be delivered simultaneously to a user even though most of them will not be utilized by the user. To improve transmission efficiency, views are usually encoded separately at the cost of coding efficiency, and transmitted on demand \cite{liu2013optimizing,toni2017optimal,toni2016network,chakareski2013user}. In this paper, we restrict our attention to MVV transmission based on separate encoding.

In \cite{toni2017optimal,toni2016network,chakareski2013user,zhang2019adaptive}, the authors consider a wired MVV system with a single server and multiple users. In particular, \cite{toni2017optimal,chakareski2013user,zhang2019adaptive} consider view synthesis only at the users, while \cite{toni2016network} considers view synthesis both at the server and users. Note that view synthesis usually introduces distortion, the degree of which depends on the distance between the synthesized view and each of its two reference views and the qualities of the two reference views. Thus, in \cite{toni2017optimal,toni2016network,chakareski2013user,zhang2019adaptive}, view selection is optimized to minimize the total distortion of all synthesized views subject to the bandwidth constraint. The transmission models in \cite{toni2017optimal,toni2016network,chakareski2013user,zhang2019adaptive} do not reflect channel fading and broadcast nature which are key features of wireless networks. Thus, the solutions for MVV transmission in \cite{toni2017optimal,toni2016network,chakareski2013user,zhang2019adaptive} cannot be directly applied to MVV transmission in multiuser wireless networks.

In \cite{zhao2015qos,zhao2014qos,zhang2018packetization,wu2015augmented}, the authors consider a wireless MVV transmission system with a single server\cite{zhao2015qos,zhao2014qos,zhang2018packetization} or multiple servers \cite{wu2015augmented} and multiple users, where channel fading and broadcast nature of wireless communications are captured. The transmission mechanisms in \cite{zhao2015qos,zhao2014qos,zhang2018packetization,wu2015augmented} make use of natural multicast opportunities to reduce energy consumption.
In particular, \cite{zhao2015qos,zhao2014qos,wu2015augmented} consider Orthogonal Frequency Division Multiple Access (OFDMA), and optimize power and subcarrier allocation to minimize the total transmission power \cite{zhao2015qos,zhao2014qos} or bandwidth consumption \cite{wu2015augmented}. 
None of \cite{zhao2015qos}, \cite{zhao2014qos} and \cite{wu2015augmented} considers view synthesis at the server or users, which can create multicast opportunities to further improve transmission efficiency in multiuser wireless networks. Thus, the transmission designs in \cite{zhao2015qos,zhao2014qos} and \cite{wu2015augmented} may be further improved. In \cite{zhang2018packetization}, the authors adopt view synthesis at each user to create multicast opportunities, but do not consider view synthesis at the server, and hence the transmission design in \cite{zhang2018packetization} cannot optimally utilize view synthesis-enabled multicast opportunities.

In this paper, we would like to address the above limitations. We consider MVV transmission from a server to multiple users in a wireless network with Time Division Multiple Access (TDMA) for multiple views. Different from \cite{zhao2015qos,zhao2014qos,wu2015augmented}, we allow view synthesis at the server and each user to maximally create multicast opportunities for efficient MVV transmission in multiuser wireless networks. The main contributions of this paper are summarized below.\footnote{This paper extends the results in the conference version \cite{xu2018energy} which does not consider the difference in timescale between view selection and time and power allocation, and studies only the energy consumption minimization problem.}

\begin{itemize}
\item  First, we establish a mathematical model to specify view synthesis at the server and each user and characterize its impact on multicast opportunities. Note that this model is highly nontrivial and fundamentally enables the optimization of multicast opportunities. To the best of our knowledge, this is the first work providing an elegant mathematical model for specifying and controlling view synthesis at the server and all users.

\item Then, we consider the optimization of view selection, transmission time and power allocation to minimize the average weighted sum energy consumption for view transmission and synthesis, for given quality requirements of all users. The problem is a challenging mixed discrete-continuous optimization problem. We propose an algorithm to obtain an optimal solution with reduced computational complexity, by exploiting optimality properties of the problem. To further reduce computational complexity, we propose a low-complexity algorithm to obtain a suboptimal solution, by transforming the original problem into a Difference of Convex (DC) problem and obtaining a stationary point of it using a DC algorithm.
\item Next, we consider the optimization of view selection, transmission time and power allocation and quality selection to maximize the total utility under the energy consumption constraints for the server and each user, respectively. The problem is more challenging, as it has extra discrete variables and the constraint functions are not tractable. By using equivalent transformations and DC programming, we propose a low-complexity algorithm to obtain a suboptimal solution.
\item  Finally, numerical results show that the proposed solutions provide substantial gains compared to existing solutions, and demonstrate the importance of the optimization of view synthesis-enabled multicast opportunities in MVV transmission.
\end{itemize}

The key notation used in this paper is listed in Table~\ref{key notations}.

\section{System Model}
\begin{figure}[!t]
	\centering
	\includegraphics[width=6.5cm]{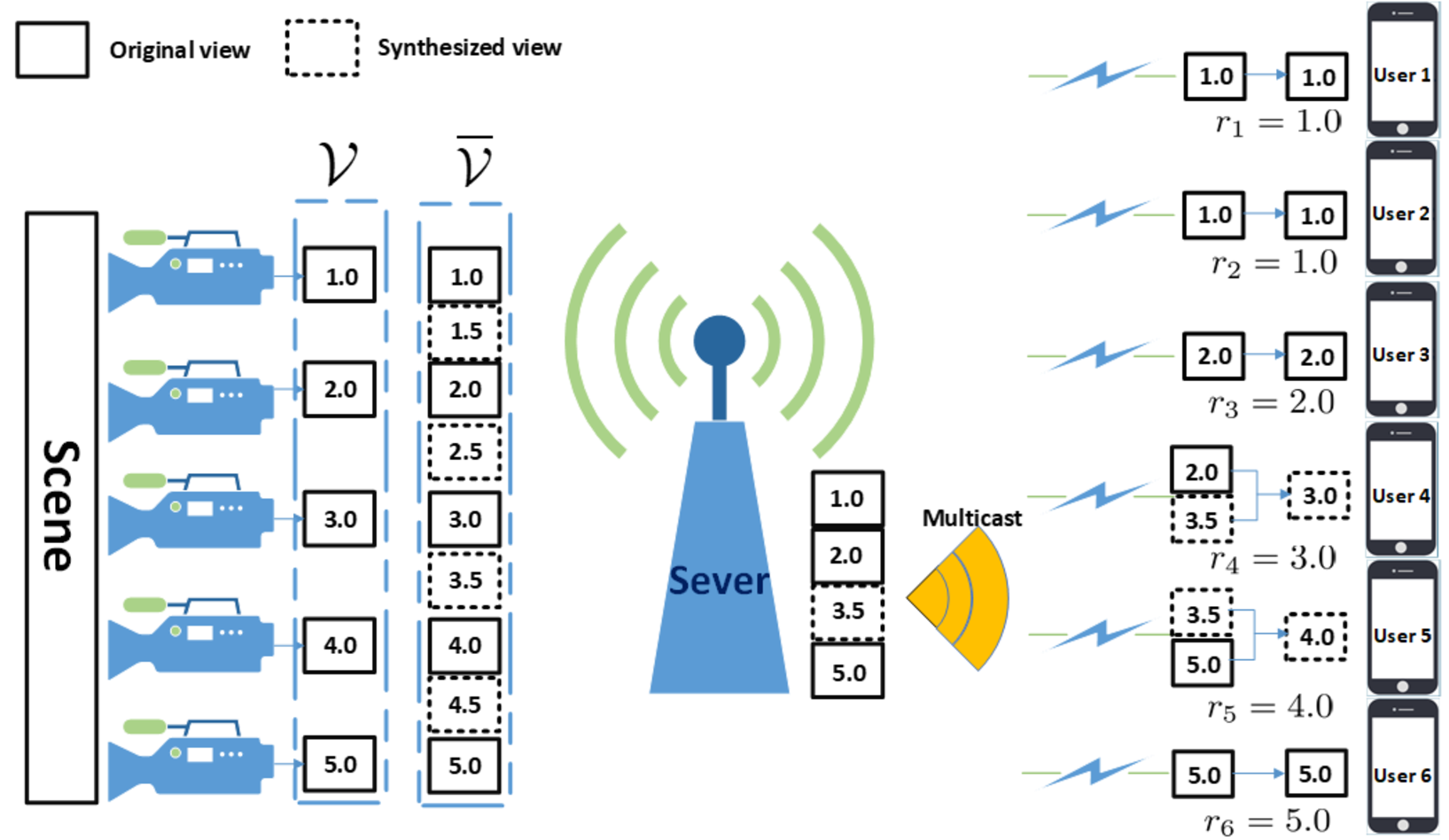}
	\caption{\small{System model. $K=6$, $r_1=1$, $r_2=1$, $r_3=2$, $r_4=3$, $r_5=4$, $r_6=5$, $V=5$,  $\mathcal{V}=\{1,2,3,4,5\}$, $\overline{\mathcal{V}}=\{1,1.5,2,\cdots,5\}$, $\Delta_k=1$ for all $k\in\mathcal{K}$, $x_1=x_2=x_{3.5}=x_{5}=1$ and $y_{1,1}=y_{2,1}=y_{3,2}=y_{4,2}=y_{4,3.5}=y_{5,3.5}=y_{5,5}=y_{6,5}=1$.}}
	\label{system model}
	\vspace{-4mm}
\end{figure}

\begin{table*}[t]
	\centering
	\caption{Key notation}
	\label{key notations}
	\begin{tabular}{|c|l|}
		\hline
		Notation & Description \\
		\hline
		$\mathcal{V}$ & set of original views \\
		\hline
		$\overline{\mathcal{V}}$ & set of all views\\
		\hline
		$\mathcal{K}$ &set of all users \\
		\hline
		$R$ & source encoding rate of all views\\
		\hline
		$\mathcal{H}$ & finite channel state space\\
		\hline
		$\Delta_k$ &  maximum allowable distance between synthesized view and  two reference views for user $k$\\
		\hline
		$r_k$ & view requested by user $k$\\
		\hline
		$x_{v}\in\{0,1\}$ & view transmission variable for view $v$\\
		\hline
		$y_{k,v}\in\{0,1\}$ & view utilization variable for view $v$ at user $k$\\
		\hline
		$t_{\mathbf{h},v}\geq 0$ & time allocated to transmit view $v$ under system channel state $\mathbf{h}$\\
		\hline
		$p_{\mathbf{h},v}\geq 0$ & power allocated to transmit view $v$ under system channel state $\mathbf{h}$\\
		\hline
		$E_b$ & synthesis energy consumption per time slot for one view at server\\
		\hline
		$E_{\text{u},k}$ & synthesis energy consumption per time slot for one view at user $k$\\
		\hline
		$R$ & source encoding rate of all views\\
		\hline
		$T$ & slot duration\\
		\hline
		$B$ & bandwidth\\
		\hline
	\end{tabular}
	\vspace{-0.3cm}
\end{table*}

As illustrated in Fig.~\ref{system model}, we consider downlink transmission of a MVV from a single-antenna server (e.g., base station or access point) to $K$ ($>$1) single-antenna users, denoted by set $\mathcal{K}\triangleq\{1,2,\cdots,K\}$. $V$ ($>$1) views (including texture maps and depth maps) about a scene of interest, denoted by set $\mathcal{V}\triangleq\{1,2,\cdots,V\}$, are captured by $V$ evenly spaced cameras simultaneously from different view angles, and are referred to as the original views. The $V$ original views are then pre-encoded independently using standard video codec and stored at the server. We consider $Q-1$ evenly spaced additional views between original view $v$ and original view $v+1$, where $Q=2,3,\cdots$ is a system parameter and $v\in\{1,2,\cdots,V-1\}$. That is, the view spacing between any two neighboring views is $1/Q$. The additional views, providing new view angles, can be synthesized via DIBR. The set of indices for all views, including the $V$ original views (which are stored at the server) and the $(V-1)(Q-1)$ additional views (which are not stored at the server but can be synthesized at the server), is denoted by $\overline{\mathcal{V}}\triangleq\{1,1+1/Q,1+2/Q,\cdots,V\}$. For ease of exposition, we assume all views have the same source encoding rate, denoted by $R$ (in bit/s).

Using  DIBR, a view can be synthesized using one left view and one right view as the reference views, at the server or a user. The quality of each synthesized view depends on its distance to its two reference views and the qualities. The server may need to synthesize any additional view $v\in\overline{\mathcal{V}}\setminus \mathcal{V}$ as it stores only the original views. Specifically, it can synthesize additional view $v\in \overline{\mathcal{V}} \setminus \mathcal{V}$ using its nearest left original view $\lfloor v \rfloor$ and right original view $\lceil v \rceil$.\footnote{$\lfloor v \rfloor$ denotes the greatest integer less than or equal to $v$, and $\lceil v \rceil$ denotes the least integer greater than or equal to $v$.} Each user $k$ may need to synthesize any view $v\in \overline{\mathcal{V}} \setminus \{1,V\}$,\footnote{Note that view $1$ and view $V$ cannot be synthesized as they are boundary views.} using two views from the left reference view set $\overline{\mathcal{V}}^-_{k,v}\triangleq\{x\in \overline{\mathcal{V}}: v-\Delta_k\leq x < v\}$ and the right reference view set $\overline{\mathcal{V}}^+_{k,v}\triangleq\{x\in\overline{\mathcal{V}}:v<x\leq v+\Delta_k\}$, respectively, where
\begin{align}
\Delta_k\in \{1,1+1/Q,\cdots,V-1\},\quad k \in \mathcal{K}. \label{Delta constraint}
\end{align}
Here, $\Delta_k$ denotes the maximum allowable distance between any synthesized view and each of its two reference views for user $k$. Thus, $\Delta_{k}$ can reflect the view quality for user $k$. That is, a smaller $\Delta_k$ indicates higher quality for user $k$.

Let $U_k(\Delta_k)$ denote the utility of user $k$ for view quality $\Delta_k$, where $U_k(\cdot)$ can be an arbitrary nonnegative, strictly decreasing and concave function\cite{6659330}. Then, for given qualities $\boldsymbol{\Delta}\triangleq (\Delta_k)_{k\in\mathcal{K}}$, the total utility is given by
\begin{equation}
U(\boldsymbol{\Delta}) \triangleq \sum_{k\in\mathcal{K}} U_k(\Delta_k).
\end{equation}

We study the system for the duration of the playback time of multiple groups of pictures (GOPs), and assume that the view angle of each user does not change within the considered duration. Note that the playback time of one GOP is usually 0.5--1 seconds. Let $r_k \in \overline{\mathcal{V}}$ denote the index of the view requested by user $k\in \mathcal{K}$. Assume $r_k,k\in\mathcal{K}$ are known at the server. To satisfy user $k$'s view request, if $r_k \in \overline{\mathcal{V}} \setminus \{1,V\}$, the server transmits either view $r_k$ or two reference views in $\overline{\mathcal{V}}^-_{k,r_k}$ and $\overline{\mathcal{V}}^+_{k,r_k}$, respectively, for user $k$ to synthesize view $r_k$; if $r_k\in\{1,V\}$, the server transmits view $r_k$. To save transmission resource by making use of multicast opportunities, the server transmits each view at most once.

Let $x_v$ denote the view transmission variable for view $v$, where
\begin{align}
x_v\in\{0,1\},\quad v\in \overline{\mathcal{V}}. \label{binary constraint x}
\end{align}
Here, $x_v=1$ indicates that the server will transmit view $v$ and $x_v=0$ otherwise. Denote $\mathbf{x} \triangleq (x_v)_{v \in \overline{\mathcal{V}}}$. As for all $v\in\overline{\mathcal{V}}\setminus\mathcal{V}$, $x_v=1$ indicates that view $v$ is synthesized at the server, $\mathbf{x}$ also reflects view synthesis at the server. Let $y_{k,v}$ denote the view utilization variable for view $v$ at user $k$, where
\begin{align}
y_{k,v}\in\{0,1\},\quad v\in \overline{\mathcal{V}},\ k \in \mathcal{K}. \label{binary constraint y}
\end{align}
Here, $y_{k,v}=1$ indicates that user $k$ will utilize view $v$ (as view $v$ is requested by user $k$, i.e., $r_k=v$, or view $v \in \overline{\mathcal{V}}^-_{k,r_k} \cup \overline{\mathcal{V}}^+_{k,r_k}$ will be used to synthesize view $r_k$ at user $k$) and $y_{k,v}=0$ otherwise. It is clear that $\mathbf{y} \triangleq \left( y_{k,v} \right)_{k\in \mathcal{K} ,v\in \overline{\mathcal{V}}}$ reflects view synthesis at all users. To guarantee that each user can obtain its requested view, we require:\footnote{For notation simplicity, for all $k\in\mathcal{K}$, we define $\sum_{v\in \overline{\mathcal{V}}^-_{k,r_k}} y_{k,v}=0$ if $ \overline{\mathcal{V}}^-_{k,r_k}=\emptyset$ and $\sum_{v\in \overline{\mathcal{V}}^+_{k,r_k}} y_{k,v}=0$ if $ \overline{\mathcal{V}}^+_{k,r_k}=\emptyset$.}
\begin{align}
&y_{k,r_k}+\sum_{\textcolor{black}{v\in \overline{\mathcal{V}}^+_{k,r_k}}} y_{k,v} = 1, \quad k\in \mathcal{K},  \label{Right constraint} \\
&y_{k,r_k}+\sum_{\textcolor{black}{v\in \overline{\mathcal{V}}^-_{k,r_k}}} y_{k,v} = 1, \quad k\in \mathcal{K}, \label{Left constraint} \\
&\sum_{\textcolor{black}{v\in \overline{\mathcal{V}} \setminus (\{r_k\}\cup \overline{\mathcal{V}}^-_{k,r_k}\cup \overline{\mathcal{V}}^+_{k,r_k})}} y_{k,v}=0,\quad k\in\mathcal{K}. \label{rest zero contraint}
\end{align}
Note that the constraints in (\ref{binary constraint y}), (\ref{Right constraint}) and (\ref{Left constraint}) \textcolor{black}{indicate that user $k$ either utilizes view $r_k$ directly, i.e.,}
\begin{align*}
&y_{k,r_k}=1,\ y_{k,v}=0, \quad \textcolor{black}{v\in \overline{\mathcal{V}}^-_{k,r_k}\cup \overline{\mathcal{V}}^+_{k,r_k}},~k\in\mathcal{K},
\end{align*}
or \textcolor{black}{utilizes one left view in $\overline{\mathcal{V}}^-_{k,r_k}$ and one right view in $\overline{\mathcal{V}}^+_{k,r_k}$ to synthesize view $r_k$, i.e.,}
\begin{align*}
&y_{k,r_k}=0,\ \sum_{v\in \textcolor{black}{\overline{\mathcal{V}}^-_{k,r_k}}}  y_{k,v}=\sum_{v\in \textcolor{black}{\overline{\mathcal{V}}^+_{k,r_k}}} y_{k,v}=1,~k\in\mathcal{K}.
\end{align*}
In addition, the constraints in (\ref{binary constraint y}) and (\ref{rest zero contraint}) \textcolor{black}{indicate that user $k$ does not utilize  view $r_k$ or views that are not in its two reference view sets, i.e.,} $$y_{k,v}=0,\quad \textcolor{black}{v\in \overline{\mathcal{V}} \setminus (\{r_k\}\cup \overline{\mathcal{V}}^-_{k,r_k}\cup \overline{\mathcal{V}}^+_{k,r_k})},~k\in\mathcal{K}.$$
The server has to transmit view $v$ in order for a user to utilize view $v$. Thus, we have the following constraints on the relation between the view transmission variables and view utilization variables:
\begin{align}
x_v \geq y_{k,v}, \quad k\in \mathcal{K},\ v\in \overline{\mathcal{V}}. \label{x>y}
\end{align}
We also refer to $(\mathbf{x},\mathbf{y})$ as view selection variables, as we can control view synthesis at the server and all users via choosing values for $(\mathbf{x},\mathbf{y})$. Due to the video coding structure, we do not allow the change of values for $(\mathbf{x},\mathbf{y})$ during the considered time duration.

The following example shows how view selection variables $(\mathbf{x},\mathbf{y})$ affect multicast opportunities.
\begin{Exam}[Natural and View Synthesis-Enabled Multicast Opportunities]
	Consider an illustration example as shown in Fig.~1. Consider $K=6$, $r_1=1$, $r_2=1$, $r_3=2$, $r_4=3$, $r_5=4$, $r_6=5$, $V=5$,  $\mathcal{V}=\{1,2,3,4,5\}$, $\overline{\mathcal{V}}=\{1,1.5,2,\cdots,5\}$ and $\Delta_k=1$ for all $k\in\mathcal{K}$. As user~1 and user~2 both request view~1, view~1 can be transmitted once to serve the two users simultaneously, corresponding to natural multicast opportunities. Without view synthesis, the server has to transmit five views, i.e., views 1, 2, 3, 4 and 5 (with view~1 being utilized by two users), making use of natural multicast opportunities. In contrast, if view synthesis is allowed at the server and each user, the server transmit only four views, i.e., views 1, 2, 3.5 and 5 (each being utilized by two users), utilizing both natural and view synthesis-enabled multicast opportunities.
\end{Exam}

Based on view selection variables $(\mathbf{x},\mathbf{y})$, the proposed model mathematically specifies view synthesis at the sever and all users, and characterizes its impact on multicast opportunities. Later, we shall see that this enables the optimization of multicast opportunities.

We consider a slotted narrowband system of bandwidth $B$ (in Hz). Consider the block fading channel model, i.e., assume the channel of each user does not change within one time slot of duration $T$ (in seconds). Note that $T$ is about 0.005 seconds. For an arbitrary time slot, let $H_k\in \mathcal{H}$ denote the random channel state of user $k$, representing the power of the channel between user~$k$ and the server, where $\mathcal{H}$ denotes the finite\footnote{\textcolor{black}{Note that we consider a finite channel state space for tractability of optimization. In addition, note that due to limited accuracy for channel estimation (and channel feedback), the operational channel state space in practical systems is finite.}} channel state space. Let $\mathbf{H} \triangleq (H_k)_{k\in\mathcal{K}}\in \mathcal{H}^K$ denote the random system channel state at an arbitrary time slot, where $\mathcal{H}^K$ represents the finite system channel state space. We assume that the server is aware of the system channel state $\mathbf{H}$ at each time slot. Suppose that the random system channel states over time slots are i.i.d. The probability of the random system channel state $\mathbf{H}$ at each time slot being $\mathbf{h}\triangleq (h_k)_{k\in\mathcal{K}} \in \mathcal{H}^K$ is given by $q_\mathbf{H}(\mathbf{h})\triangleq\Pr[\mathbf{H}=\mathbf{h}]$.

We consider Time Division Multiple Access (TDMA)\footnote{Note that TDMA is analytically tractable and has applications in WiFi systems. In addition, the proposed transmission scheme and the optimization framework for a TDMA system can be extended to an OFDMA system~\cite{guo2018optimal}.} \textcolor{black}{for multiple views. That is, different views are transmitted one after another over the same frequency channel}. Consider an arbitrary time slot. The time allocated to transmit view $v$ under the system channel state $\mathbf{h}$, denoted by $t_{\mathbf{h},v}$, satisfies:
\begin{equation}
t_{\mathbf{h},v} \geq 0,\quad  \mathbf{h}\in\mathcal{H}^K,\ v\in \overline{\mathcal{V}}. \label{t>=0}
\end{equation}
In addition, we have the following total time allocation constraints under the system channel state $\mathbf{h}$:
\begin{equation}
\sum_{v\in \overline{\mathcal{V}}} t_{\mathbf{h},v} \leq T,\quad \mathbf{h}\in\mathcal{H}^K. \label{time constraint}
\end{equation}

The transmission power for view $v$ under the system channel state $\mathbf{h}$, denoted by $p_{\mathbf{h},v}$, satisfies:
\begin{equation}
p_{\mathbf{h},v} \geq 0,\quad  \mathbf{h}\in\mathcal{H}^K,\ v\in \overline{\mathcal{V}}. \label{p>=0}
\end{equation}
The maximum transmission rate of view $v$ to user $k$ \blue{under the system channel state $\mathbf{h}$} is given by \blue{$\frac{Bt_{\mathbf{h},v}}{T} \log_2\left(1+\frac{p_{\mathbf{h},v} h_k}{\sigma^2} \right)$} (in bits/s), where $\blue{\sigma^2}$ is the power \textcolor{black}{(in Watt)} of the complex additive white Gaussian channel noise at each receiver. To \textcolor{black}{reduce the chance of stall (i.e., the chance that a playback buffer is empty)} during the video playback at each user, \blue{the average arrival rate of each playback queue should be no less than its service rate. Thus,} we have the following successful transmission constraints:\footnote{\textcolor{black}{More conservatively, we can use $R+\delta$ for some $\delta>0$ in (\ref{bandwidth constraint}) instead of $R$. In addition, in the startup phase, the playback buffer usually stores some view data, say $L_0$ bits. It is known that the chance of stall during the playback phase decreases with $\delta$ and with $L_0$.}}
\begin{equation}
\frac{B}{T}\textcolor{black}{\mathbb{E}_{\mathbf{H}}} \left[ t_{\mathbf{H},v} \log_2\left(1+\frac{p_{\mathbf{H},v} H_k}{\blue{\sigma^2}}\right) \right] \geq  y_{k,v} R, \quad k\in \mathcal{K},\  v\in \overline{\mathcal{V}}, \label{bandwidth constraint}
\end{equation}
\blue{where the expectation $\mathbb{E}_{\mathbf{H}}$ is taken over the random system channel state $\mathbf{H} \in \mathcal{H}^K$.} The transmission energy consumption per time slot under the system channel state $\mathbf{h}$ at the server is
$\sum_{v\in \overline{\mathcal{V}}} t_{\mathbf{h},v}p_{\mathbf{h},v}$. Besides view transmission, view synthesis also consumes energy. For ease of exposition, we assume that the energy consumptions for synthesizing different views at the server are the same. Let $E_b$ denote the synthesis energy consumption \textcolor{black}{(in Joule)}\cite{6195536} per time slot for one view at the server. Thus, the total synthesis energy consumption per time slot at the server is
$\sum_{v\in \overline{\mathcal{V}} \setminus \mathcal{V}}x_vE_b \label{synthesis energy at server}$.
Let $E_{\text{u},k}$ denote the synthesis energy consumption \textcolor{black}{(in Joule)} per time slot for one view at user $k$. Considering heterogeneous hardware conditions at different users, we allow $E_{\text{u},k},k\in \mathcal{K}$ to be different. Then, the synthesis energy consumption per time slot at user $k$ is $(1-y_{k,r_k})E_{\text{u},k}$, and the total synthesis energy consumption per time slot at all users is
$\sum_{k\in \mathcal{K}} (1-y_{k,r_k})E_{\text{u},k}$.
Therefore, the weighted sum energy consumption per time slot under the system channel state $\mathbf{h}$ is given by:
\begin{align}
E(\mathbf{x},\mathbf{y},\mathbf{t}_\mathbf{h},\mathbf{p}_\mathbf{h}) &
= \sum_{v\in \overline{\mathcal{V}}} t_{\mathbf{h},v}p_{\mathbf{h},v}+\sum_{v\in \overline{\mathcal{V}} \setminus \mathcal{V}}x_vE_b\nonumber\\
&+\beta \sum_{k\in \mathcal{K}} (1-y_{k,r_k})E_{\text{u},k},\quad \mathbf{h}\in\mathcal{H}^K,
\end{align}
where $\mathbf{x}\triangleq (x_v)_{v\in \overline{\mathcal{V}}}$, $\mathbf{y}\triangleq (y_{k,v})_{k\in\mathcal{K},v\in \overline{\mathcal{V}}}$, $\mathbf{t}_\mathbf{h}\triangleq (t_{\mathbf{h},v})_{v\in \overline{\mathcal{V}}}$, $\mathbf{p}_\mathbf{h}\triangleq (p_{\mathbf{h},v})_{v\in \overline{\mathcal{V}}}$ and $\beta\geq 1$ is the corresponding weight factor for the $K$ users. Note that $\beta >1 $ means imposing a higher cost on the energy consumptions for user devices due to their limited battery powers. The average weighted sum energy consumption per time slot is given by
\begin{align}
\mathbb{E}_{\mathbf{H}}[E(\mathbf{x},\mathbf{y},\mathbf{t}_\mathbf{H},\mathbf{p}_\mathbf{H})] &= \textcolor{black}{\mathbb{E}_{\mathbf{H}}} \left[\sum_{v\in \overline{\mathcal{V}}} t_{\mathbf{H},v}p_{\mathbf{H},v}\right]+\sum_{v\in \overline{\mathcal{V}} \setminus \mathcal{V}}x_vE_b\nonumber \\
&+\beta \sum_{k\in \mathcal{K}} (1-y_{k,r_k})E_{\text{u},k}. \label{expected energy objective}
\end{align}

In Section~\ref{Part I}, we \textcolor{black}{shall} minimize the average weighted sum energy consumption for given quality requirements of all users. In Section~\ref{Part II}, we \textcolor{black}{shall} maximize the total utility under the energy consumption constraints at the server and each user.
\section{Average Weighted Sum Energy Minimization}\label{Part I}
In this section, we consider the minimization of the average weighted sum energy consumption for given quality requirements of all users. We first formulate the optimization problem. Then, we develop an algorithm to obtain an optimal solution with reduced computational complexity by exploiting optimality properties. Finally, to further reduce computational complexity, we develop a low-complexity algorithm to obtain a suboptimal solution using DC programming.
\subsection{Problem Formulation}
We would like to minimize the average weighted sum energy consumption by optimizing the view selection and transmission time and power allocation for given quality requirements of all users. Specifically, for given $\boldsymbol{\Delta}$, we have the following optimization problem.
\begin{problem}[Energy Minimization]\label{View selection and resource allocation}
\begin{align}
	E^\star \triangleq &\min_{\mathbf{x},\mathbf{y},\mathbf{t},\mathbf{p}}\quad \textcolor{black}{\mathbb{E}_{\mathbf{H}}}\left[E(\mathbf{x},\mathbf{y},\mathbf{t}_\mathbf{H},\mathbf{p}_\mathbf{H})\right] \notag\\
	&~~\text{s.t.} \quad (\ref{binary constraint x}),(\ref{binary constraint y}),(\ref{Right constraint}),(\ref{Left constraint}),(\ref{rest zero contraint}),(\ref{x>y}),(\ref{t>=0}),(\ref{time constraint}),(\ref{p>=0}),(\ref{bandwidth constraint}), \notag
\end{align}
\end{problem}
where $\textcolor{black}{\mathbb{E}_{\mathbf{H}}}[E(\mathbf{x},\mathbf{y},\mathbf{t}_\mathbf{H},\mathbf{p}_\mathbf{H})]$ is given by (\ref{expected energy objective}). Let ($\mathbf{x}^\star,\mathbf{y}^\star,\mathbf{t}^\star,\mathbf{p}^\star$) denote an optimal solution of Problem \ref{View selection and resource allocation}, where $\mathbf{x}^\star \triangleq (x_v^\star)_{v\in \overline{\mathcal{V}}}$, $\mathbf{y}^\star \triangleq (y^\star_{k,v})_{k\in \mathcal{K},v\in \overline{\mathcal{V}}}$, $\mathbf{t}^\star \triangleq (t^\star_{\mathbf{h},v})_{\mathbf{h} \in \mathcal{H}^K , v\in \overline{\mathcal{V}}}$ and  $\mathbf{p}^\star \triangleq (p^\star_{\mathbf{h},v})_{\mathbf{h} \in \mathcal{H}^K, v\in \overline{\mathcal{V}}}$.

Problem \ref{View selection and resource allocation} is a challenging mixed discrete-continuous optimization problem with two types of variables, i.e., binary view selection variables $(\mathbf{x},\mathbf{y})$ as well as continuous power allocation and time allocation variables $(\mathbf{t},\mathbf{p})$. For given $(\mathbf{x},\mathbf{y})$, the optimization with respect to $(\mathbf{t},\mathbf{p})$ is nonconvex,  as $ \sum_{v\in \overline{\mathcal{V}}} t_{\mathbf{h},v}p_{\mathbf{h},v}$ is not convex in $(\mathbf{t}_\mathbf{h},\mathbf{p}_\mathbf{h})$.
\subsection{Optimal Solution}\label{dual decomposition}
In this part, we develop an algorithm to obtain an optimal solution of Problem~\ref{View selection and resource allocation}. Define $\mathbf{Y} \triangleq \{ \mathbf{y}: (\ref{binary constraint y}),(\ref{Right constraint}),(\ref{Left constraint}),(\ref{rest zero contraint}) \}$ and $\mathbf{X}\times \mathbf{Y} \triangleq \{(\mathbf{x},\mathbf{y}): (\ref{binary constraint x}),(\ref{x>y}), \mathbf{y}\in \mathbf{Y}\}$. First, by a change of variables, i.e., using $e_{\mathbf{h},v}\triangleq p_{\mathbf{h},v} t_{\mathbf{h},v}$ (representing the transmission energy for view $v$ under $\mathbf{h}$) instead of $p_{\mathbf{h},v}$ for all $\mathbf{h}\in \mathcal{H}^K,v\in\overline{\mathcal{V}}$, and by exploiting structural properties of Problem~\ref{View selection and resource allocation}, we obtain an equivalent problem of Problem~\ref{View selection and resource allocation}.
\begin{problem}[View Selection]\label{View selection}
\begin{align}
    &\min_{(\mathbf{x},\mathbf{y}) \in \mathbf{X}\times \mathbf{Y}} E_\text{t}^\star(\mathbf{y})+\sum_{v\in \overline{\mathcal{V}} \setminus \mathcal{V}}x_vE_b+\beta \sum_{k\in \mathcal{K}} (1-y_{k,r_k})E_{\text{u},k} \notag
\end{align}
\end{problem}
where $E_\text{t}^\star(\mathbf{y})$ is given by the following problem. Let $(\mathbf{x}^\star, \mathbf{y}^\star)$ denote an optimal solution of Problem~\ref{View selection}.
\begin{problem}[Time and Energy Allocation for Given $\mathbf{y}$]\label{sub-problem for resource allocation} For any given $\mathbf{y}\in\mathbf{Y}$,
	\begin{align}
	E_\text{t}^\star(\mathbf{y})& \triangleq\min_{\mathbf{t},\mathbf{e}} \quad \textcolor{black}{\mathbb{E}_{\mathbf{H}}} \left[ \sum_{v \in \overline{\mathcal{V}}} e_{\mathbf{H},v}\right] \notag\\
	&~\text{s.t.}  \quad  \textcolor{black}{(\ref{t>=0}),(\ref{time constraint}),} \notag\\
	&~ \quad \quad e_{\mathbf{h},v}\geq 0,\quad \mathbf{h}\in \mathcal{H}^K,\ v\in\overline{\mathcal{V}}, \label{e>=0} \\
	&~ \quad \quad \frac{B}{T} \textcolor{black}{\mathbb{E}_{\mathbf{H}}} \left[ t_{\mathbf{H},v} \log_2\left(1+\frac{e_{\mathbf{H},v} H_k}{t_{\mathbf{H},v}\blue{\sigma^2}}\right) \right] \geq  y_{k,v} R,\nonumber\\
	&\qquad \qquad\qquad\qquad\qquad\qquad k\in \mathcal{K},\  v\in \overline{\mathcal{V}}.\label{transformed bandwidth constraint}
	\end{align}
\end{problem}
Let $(\mathbf{t}^\star(\mathbf{y}),\mathbf{e}^\star(\mathbf{y}))$ denote an optimal solution of Problem~\ref{sub-problem for resource allocation}, where $\mathbf{t}^\star(\mathbf{y})\triangleq (t^\star_{\mathbf{h},v} \left(\mathbf{y})\right)_{\mathbf{h}\in \mathcal{H}^K,v\in\overline{\mathcal{V}}}$ and $\mathbf{e}^\star(\mathbf{y})\triangleq (e^\star_{\mathbf{h},v} \left(\mathbf{y})\right)_{\mathbf{h}\in \mathcal{H}^K,v\in\overline{\mathcal{V}}}$.

Note that the constraints in (\ref{p>=0}) and (\ref{bandwidth constraint}) are equivalent to the constraints in (\ref{e>=0}) and (\ref{transformed bandwidth constraint}), respectively. This formulation (including Problem~\ref{View selection} and Problem~\ref{sub-problem for resource allocation}) separates the two types of variables (i.e., binary variables and continuous variables) and facilitates the optimization. Due to the equivalence between Problem~\ref{View selection and resource allocation} and Problems~\ref{View selection} and \ref{sub-problem for resource allocation}, we know that $(\mathbf{x}^\star,\mathbf{y}^\star,\mathbf{t}^\star(\mathbf{y}^\star),\mathbf{p}^\star(\mathbf{y}^\star))$ is an optimal solution of Problem~\ref{View selection and resource allocation}, where $\mathbf{p}^\star(\mathbf{y})\triangleq (p^\star_{\mathbf{h},v} \left(\mathbf{y})\right)_{\mathbf{h}\in \mathcal{H}^K,v\in\overline{\mathcal{V}}}$ with $p^\star_{\mathbf{h},v}(\mathbf{y}) = e^\star_{\mathbf{h},v}(\mathbf{y}) / t^\star_{\mathbf{h},v}(\mathbf{y})$, $\mathbf{h}\in \mathcal{H}^K,v\in\overline{\mathcal{V}}$. Thus, we can obtain an optimal solution of Problem~\ref{View selection and resource allocation} by first solving Problem~\ref{sub-problem for resource allocation} and then solving Problem~\ref{View selection}.

\subsubsection{Solution of Problem~3}\label{solution of p3}
First, we focus on solving Problem~\ref{sub-problem for resource allocation}. Problem~\ref{sub-problem for resource allocation} is a convex optimization problem and can be solved using standard convex optimization techniques \cite{boyd2004convex}.\footnote{\textcolor{black}{In this paper, we assume that the server is aware of the statistics of the random system channel. Thus, the problems considered in this paper are not stochastic optimization problems.}} Note that when the \textcolor{black}{sizes} of $\mathcal{H}$ and $K$ are large, the numbers of variables and constraints in Problem~\ref{sub-problem for resource allocation} are huge, leading to high computational complexity. As Problem~\ref{sub-problem for resource allocation} is convex and strictly feasible, implying that Slater's condition holds, the duality gap is zero. To accelerate the speed for solving Problem~\ref{sub-problem for resource allocation}, we can also adopt partial dual decomposition to enable parallel computation \cite{boyd2007notes}. Specifically, by relaxing the coupling constraints in (\ref{transformed bandwidth constraint}), we obtain a decomposable partial dual problem of Problem~\ref{sub-problem for resource allocation}.
\begin{problem}[Partial Dual Problem of Problem~\ref{sub-problem for resource allocation}]\label{master problem} For any given $\mathbf{y}\in\mathbf{Y}$,
	\begin{align}
	D^\star(\mathbf{y})\triangleq&\max_{\boldsymbol{\lambda}}  \quad D(\mathbf{y},\boldsymbol{\lambda})=\sum_{\mathbf{h}\in \mathcal{H}^K} q_\mathbf{H}(\mathbf{h})D_\mathbf{h}(\mathbf{y},\boldsymbol{\lambda}) \notag\\
	&~\text{s.t.} \quad \lambda_{k,v}\geq 0,\quad k \in \mathcal{K},\ v\in\overline{\mathcal{V}},
	\end{align}
\end{problem}
where $\boldsymbol{\lambda}\triangleq(\lambda_{k,v})_{k\in\mathcal{K},v\in\overline{\mathcal{V}}}$. Let $\boldsymbol{\lambda}^\star(\mathbf{y})$ denote an optimal solution of Problem~\ref{master problem}. $D_\mathbf{h}(\mathbf{y},\boldsymbol{\lambda})$ is given by the following subproblem.
\begin{problem}[Subproblem of Problem~\ref{master problem} for $\mathbf{h}\in \mathcal{H}^K$]\label{decompostion} For any given $\mathbf{y}\in\mathbf{Y}$ and $\boldsymbol{\lambda}\succeq0$,
	\begin{align}
	&D_\mathbf{h}(\mathbf{y},\boldsymbol{\lambda})\triangleq\nonumber\min_{\mathbf{t}_{\mathbf{h}},\mathbf{e}_{\mathbf{h}}}\quad  \sum_{v \in \overline{\mathcal{V}}} e_{\mathbf{h},v}\\
	& \qquad - \sum_{k\in \mathcal{K}} \sum_{v \in \overline{\mathcal{V}}} \lambda_{k,v} \left(  \frac{B}{T} t_{\mathbf{h},v} \log_2\left(1+\frac{e_{\mathbf{h},v} h_k}{t_{\mathbf{h},v}\blue{\sigma^2}}\right) -y_{k,v}R \right) \notag\\
	&~\text{s.t.} \quad e_{\mathbf{h},v}\geq 0,\quad v\in\overline{\mathcal{V}},\\
	&~ \quad \quad  t_{\mathbf{h},v}\geq 0,\quad v\in\overline{\mathcal{V}},\\
	&~ \quad \quad  \sum_{v\in \overline{\mathcal{V}}}t_{\mathbf{h},v}\leq T,
	\end{align}
\end{problem}
where $\mathbf{e}_{\mathbf{h}}\triangleq(e_{\mathbf{h},v})_{v\in\overline{\mathcal{V}}}$. Let $\left(\mathbf{t}_{\mathbf{h}}^\star(\mathbf{y},\boldsymbol{\lambda}),\mathbf{e}^\star_{\mathbf{h}}(\mathbf{y},\boldsymbol{\lambda})\right)$ denote an optimal solution of Problem~\ref{decompostion}, where $\mathbf{t}_{\mathbf{h}}^\star(\mathbf{y},\boldsymbol{\lambda}) \triangleq (t_{\mathbf{h},v}^\star(\mathbf{y},\boldsymbol{\lambda}))_{v\in\overline{\mathcal{V}}}$ and $\mathbf{e}_{\mathbf{h}}^\star(\mathbf{y},\boldsymbol{\lambda}) \triangleq (e_{\mathbf{h},v}^\star(\mathbf{y},\boldsymbol{\lambda}))_{v\in\overline{\mathcal{V}}}$.

Define $\mathbf{t}^\star(\mathbf{y},\boldsymbol{\lambda}) \triangleq (\mathbf{t}_{\mathbf{h}}^\star(\mathbf{y},\boldsymbol{\lambda}))_{\mathbf{h}\in \mathcal{H}^K}$ and $\mathbf{e}^\star(\mathbf{y},\boldsymbol{\lambda}) \triangleq (\mathbf{e}_{\mathbf{h}}^\star(\mathbf{y},\boldsymbol{\lambda}))_{\mathbf{h}\in \mathcal{H}^K}$. We have the following result.

\begin{lemma}[Relationship between Problems~\ref{master problem},\ref{decompostion} and Problem~\ref{sub-problem for resource allocation}] \label{dual property}
For any given $\mathbf{y}\in\mathbf{Y}$, $D^\star(\mathbf{y})=E_\text{t}^\star(\mathbf{y})$,  $\mathbf{t}^\star(\mathbf{y})=\mathbf{t}^\star(\mathbf{y},\boldsymbol{\lambda}^\star(\mathbf{y}))$ and $\mathbf{e}^\star (\mathbf{y})=\mathbf{e}^\star (\mathbf{y},\boldsymbol{\lambda}^\star(\mathbf{y}))$.
	\begin{IEEEproof}
		Please refer to \textcolor{black}{Appendix A}.
	\end{IEEEproof}
\end{lemma}

By Lemma~\ref{dual property}, we can obtain an optimal solution of Problem~\ref{sub-problem for resource allocation} by solving Problem~\ref{master problem} and Problem~\ref{decompostion} for all $\mathbf{h}\in \mathcal{H}^K$. As Problem~\ref{decompostion} for all $\mathbf{h}\in\mathcal{H}^K$ can be solved in parallel using standard convex optimization techniques, we can compute $(\mathbf{t}^\star(\mathbf{y},\boldsymbol{\lambda}),\mathbf{e}^\star(\mathbf{y},\boldsymbol{\lambda}))$ efficiently. In addition, Problem~\ref{master problem} is convex and can be solved using the subgradient method~\cite{bertsekas1999nonlinear}. In particular, for all $k\in\mathcal{K}, v\in\overline{\mathcal{V}}$, the subgradient method generates a sequence of dual feasible points according to the following update equation:
\begin{equation}
\lambda_{k,v}{(n+1)} = \max\{\lambda_{k,v}{(n)}+\eta_{k,v}(n)s_{k,v}(\mathbf{y},\boldsymbol{\lambda}(n)),0\},\label{update lambda}
\end{equation}
where $\boldsymbol{\lambda}(n)\triangleq(\lambda_{k,v}(n))_{k\in\mathcal{K},v\in\overline{\mathcal{V}}}$ and $s_{k,v}(\mathbf{y},\boldsymbol{\lambda}(n))$ denotes a subgradient of $D(\mathbf{y},\boldsymbol{\lambda}(n))$ with respect to $\lambda_{k,v}$  given by:
\begin{align}
&s_{k,v}(\mathbf{y},\boldsymbol{\lambda}(n))\triangleq y_{k,v}R \nonumber \\
&-\frac{B}{T}\sum_{\mathbf{h}\in \mathcal{H}^K}   q_\mathbf{H}(\mathbf{h}) t_{\mathbf{h},v}^\star(\mathbf{y},\boldsymbol{\lambda}(n)) \log_2\left(1+\frac{e_{\mathbf{h},v}^\star(\mathbf{y},\boldsymbol{\lambda}(n)) h_k}{t_{\mathbf{h},v}^\star(\mathbf{y},\boldsymbol{\lambda}(n)) \blue{\sigma^2}}\right). \label{subgradient}
\end{align}
Here, $n$ denotes the iteration index and $\{\eta_{k,v}(n)\}$ is a step size sequence, satisfying\textcolor{black}{:}
\begin{align*}
&\eta_{k,v}(n)> 0,~\sum_{n=0}^{\infty} \eta_{k,v}(n) = \infty,\\ &\sum_{n=0}^{\infty} \eta_{k,v}^2(n) < \infty,~ \lim_{n\rightarrow \infty} \eta_{k,v}(n)=0.
\end{align*}
It has been shown in \cite{bertsekas1999nonlinear} that $\boldsymbol{\lambda}(n)\rightarrow \boldsymbol{\lambda}^\star(\mathbf{y})$ as $n \rightarrow \infty$, for all initial points $\boldsymbol{\lambda}(0) \succeq 0$. Therefore, we can solve Problem~\ref{sub-problem for resource allocation} by solving Problem~\ref{master problem} and Problem~\ref{decompostion} for all $\mathbf{h}\in\mathcal{H}^K$.
\subsubsection{Solution of Problem~\ref{View selection}}\label{solution of p2}
Next, we focus on solving Problem~\ref{View selection}, which is a challenging discrete optimization problem. Problem~\ref{View selection} can be solved by exhaustive search over $\mathbf{X}\times\mathbf{Y}$. \textcolor{black}{We would like to reduce the search space by analyzing optimality properties of Problem~\ref{View selection}}. For any two users $a \in \mathcal{K}$ and $b \in \mathcal{K}$, define $r_{\max} \triangleq\max \{r_a,r_b\}$, $r_{\min} \triangleq\min \{r_a,r_b\}$ and $\mathcal{L}_{a,b}$ is given by (\ref{lem2:lab}), as shown at the top of the next page.
\begin{figure*}
	
\begin{align}
&\mathcal{L}_{a,b} \triangleq \begin{cases}
\{r_a\}, & \overline{\mathcal{V}}_{r_{\min}}^+ \cap \overline{\mathcal{V}}_{r_{\max}}^- = \emptyset,\\
\{ r_a,r_{\max}-\Delta,r_{\min}+\Delta\}\cup (\overline{\mathcal{V}}_{r_{\min}}^+ \cap \overline{\mathcal{V}}_{r_{\max}}^- \cap \mathcal{V}), & \overline{\mathcal{V}}_{r_{\min}}^+ \cap \overline{\mathcal{V}}_{r_{\max}}^- \neq \emptyset\text{ and } r_{\max} \notin \overline{\mathcal{V}}_{r_{\min}}^+, \\
\{ r_a,r_b,r_{\max}-\Delta,r_{\min}+\Delta\},& r_{\max} \in \overline{\mathcal{V}}_{r_{\min}}^+.
\end{cases}\label{lem2:lab}
\end{align}
\normalsize \hrulefill
\end{figure*}
For all user $k\in\mathcal{K}$, define $\mathcal{L}_k \triangleq \bigcup_{i \in \mathcal{K}:i\neq k} \mathcal{L}_{k,i}.$ We have the following lemma.
\begin{lemma}[Optimality Properties of Problem~\ref{View selection}] \label{optimality properties}
(i) $x_v^\star = \max_{k\in \mathcal{K}} y_{k,v}^\star$, $ v \in \overline{\mathcal{V}}$;
(ii) Suppose $\beta E_{\text{u},k}\geq E_{\text{b}}$, $k \in \mathcal{K}$. \textcolor{black}{Then, $y^\star_{k,v}=0$, $k\in \mathcal{K}$, $v \in \overline{\mathcal{V}} \setminus \left(\cup_{k\in\mathcal{K}} \mathcal{L}_k\right)$.}
\begin{IEEEproof}
	Please refer to Appendix B.
\end{IEEEproof}
\end{lemma}
 \begin{figure}[t]
	\subfigure[$\overline{\mathcal{V}}_{r_{\min}}^+ \cap \overline{\mathcal{V}}_{r_{\max}}^- = \emptyset$]{
		\begin{minipage}{\linewidth}	
			\centering
			\includegraphics[width=7cm]{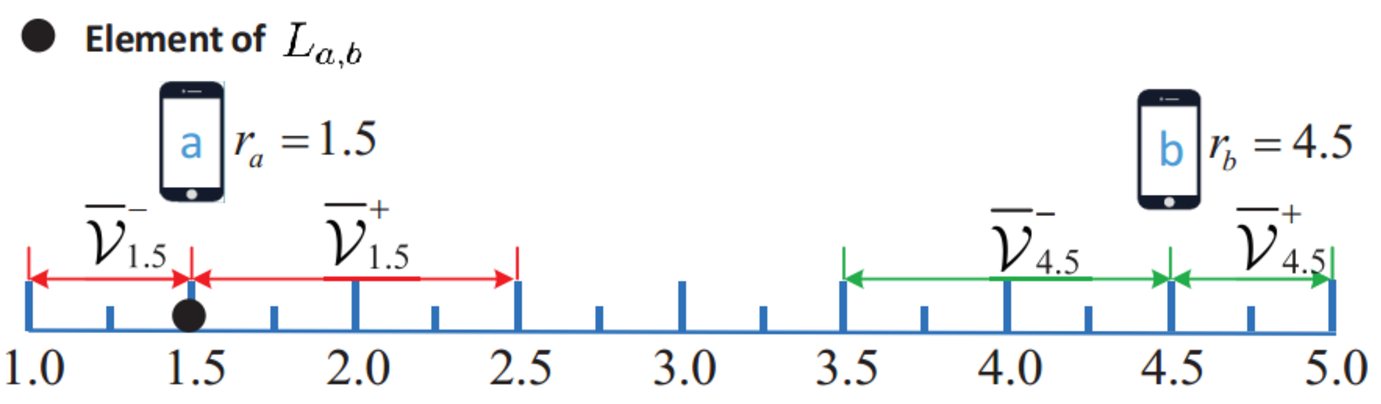}
		\end{minipage}
	}
	\subfigure[$\overline{\mathcal{V}}_{r_{\min}}^+ \cap \overline{\mathcal{V}}_{r_{\max}}^- \neq \emptyset$ and $r_{\max} \notin \overline{\mathcal{V}}_{r_{\min}}^+$]{
		\begin{minipage}{\linewidth}
			\centering
			\includegraphics[width=7cm]{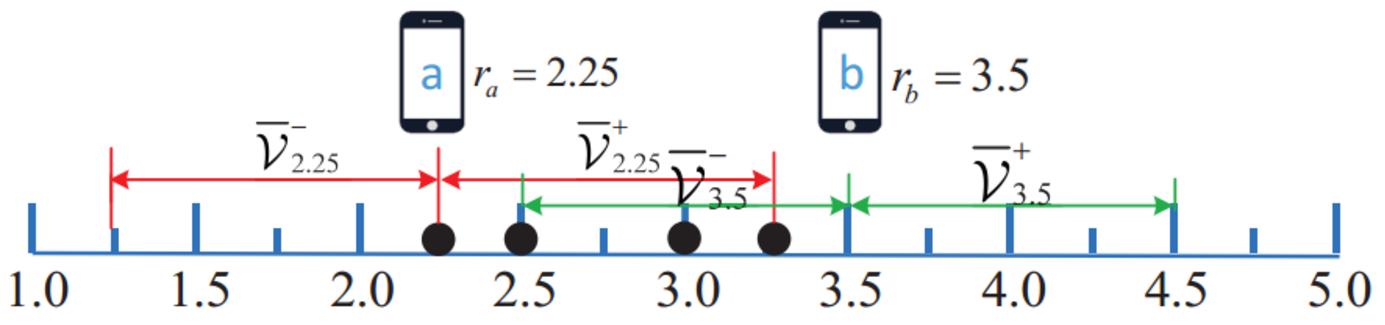}
		\end{minipage}
	}
	\subfigure[$ r_{\max} \in \overline{\mathcal{V}}_{r_{\min}}^+$]{
		\begin{minipage}{\linewidth}
			\centering
			\includegraphics[width=7cm]{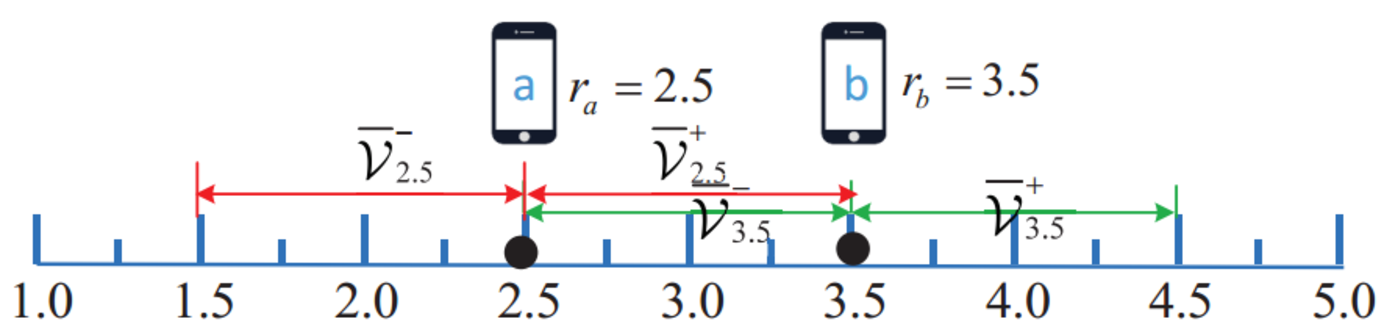}
		\end{minipage}
	}
	\caption{\textcolor{black}{Illustration of $\mathcal{L}_{a,b}$. $\mathcal{V}=\{1,2,3,4,5\}$, $\overline{\mathcal{V}}=\{1,1.25,1.5,\cdots,5\}$ and $\Delta_k=1,k\in\mathcal{K}$.}}
	\label{Illustration}
\end{figure}

Statement~(i) indicates that view $v$ will be transmitted \textcolor{black}{if at least one user} utilizes it. \textcolor{black}{$\mathcal{L}_{a,b}$ \textcolor{black}{can be viewed as} the set of views that may be utilized by user $a$ when considering the presence of only users $a$ and $b$, as illustrated in Fig.~\ref{Illustration}; $\mathcal{L}_k$ \textcolor{black}{can be interpreted as} the set of views that may be utilized by user $k$ considering the presence of all users; $\cup_{k\in\mathcal{K}} \mathcal{L}_k$ \textcolor{black}{can be treated as} the set of views that may be utilized by at least one user. Thus, Statement~(ii) indicates that no views in $\overline{\mathcal{V}} \setminus \left(\cup_{k\in\mathcal{K}} L_k\right)$ will be utilized by any user.} \textcolor{black}{Lemma~\ref{optimality properties} characterizes the relationship between $\mathbf{x}^\star$ and $\mathbf{y}^\star$, and determines some zero elements of $\mathbf{y}^\star$}. Let $\underline{\mathbf{X}} \times \underline{\mathbf{Y}} \triangleq \{ (\mathbf{x},\mathbf{y}) |\ \mathbf{y}\in \underline{\mathbf{Y}},x_v= \max_{k\in\mathcal{K}} y_{k,v}, v\in \overline{\mathcal{V}}  \}$, where $\underline{\mathbf{Y}} \triangleq \{\mathbf{y} \in \mathbf{Y}|y_{k,v}=0, v \in  \overline{\mathcal{V}} \setminus \left(\cup_{k\in\mathcal{K}} U_k\right), k\in \mathcal{K}\}$. Based on Lemma~\ref{optimality properties}, we can reduce the feasible set for $(x,y)$ from $\mathbf{X} \times \mathbf{Y}$ to $\underline{\mathbf{X}} \times \underline{\mathbf{Y}}$ without losing optimality.
\subsubsection{Algorithm}
Based on the \textcolor{black}{results in Section~\ref{solution of p3} and Section~\ref{solution of p2}}, we develop an algorithm to obtain an optimal solution of Problem~\ref{View selection and resource allocation}, as summarized in Algorithm~1.
\begin{algorithm}[t]
	\small
	\caption{Algorithm for Obtaining \textcolor{black}{an} Optimal Solution of Problem~\ref{View selection and resource allocation}}
	\textbf{Output} ($\mathbf{x}^\star$,$\mathbf{y}^\star$,$\mathbf{t}^\star$,$\mathbf{p}^\star$).
	\begin{algorithmic}[1]
	\STATE Set $E^\star=\infty$, $n=0$, and choose any $\boldsymbol{\lambda}(n)\geq 0$.
	\FOR{$(\mathbf{x},\mathbf{y}) \in \underline{\mathbf{X}} \times \underline{\mathbf{Y}}$}
	\REPEAT
	\STATE For all $\mathbf{h} \in \mathcal{H}^K$, obtain $(\mathbf{t}_{\mathbf{h}}(\mathbf{y},\boldsymbol{\lambda}(n)),\mathbf{e}_{\mathbf{h}}(\mathbf{y},\boldsymbol{\lambda}(n)))$ by solving Problem~\ref{decompostion} using standard convex optimization techniques.
	\STATE For all $k\in\mathcal{K}$ and $v\in\overline{\mathcal{V}}$, compute $\lambda_{k,v}$ according to (\ref{update lambda}), where $s_{k,v}(\mathbf{y},\boldsymbol{\lambda}(n))$ is obtained according to (\ref{subgradient}).
	\STATE Set $n=n+1$.
	\UNTIL convergence criteria is met.
	\STATE For all $\mathbf{h}\in \mathcal{H}^K$, set $\mathbf{t}_{\mathbf{h}}(\mathbf{y})=\mathbf{t}_{\mathbf{h}}(\mathbf{y},\boldsymbol{\lambda}(n-1))$ and $\mathbf{e}_{\mathbf{h}}(\mathbf{y})=\mathbf{e}_{\mathbf{h}}(\mathbf{y},\boldsymbol{\lambda}(n-1))$.
	\STATE Compute $E(\mathbf{x},\mathbf{y})=\sum_{\mathbf{h}\in\mathcal{H}^K}q_{\mathbf{H}}(\mathbf{h})\mathbf{e}_{\mathbf{h}}(\mathbf{y})+\sum_{v\in \overline{\mathcal{V}} \setminus \mathcal{V}}x_vE_b+\beta \sum_{k\in \mathcal{K}} (1-y_{k,r_k})E_{\text{u},k}$.
	\IF{$E(\mathbf{x},\mathbf{y}) \leq E^\star$}
	\STATE Obtain $\mathbf{p}=(p_{\mathbf{h},v})_{\mathbf{h}\in\mathcal{H}^K,v\in\overline{\mathcal{V}}}$ where $p_{\mathbf{h},v}=e_{\mathbf{h},v}/t_{\mathbf{h},v},\mathbf{h}\in \mathcal{H}^K$, $v\in \overline{\mathcal{V}}$.
	\STATE Set $E^\star = E(\mathbf{x},\mathbf{y})$ and  $(\mathbf{x}^\star,\mathbf{y}^\star,\mathbf{t}^\star,\mathbf{p}^\star) =( \mathbf{x},\mathbf{y},\mathbf{t},\mathbf{p})$.
	\ENDIF
	\ENDFOR
	\end{algorithmic}
\end{algorithm}

\subsection{Suboptimal Solution}
Although the complexity for obtaining an optimal solution of Problem~\ref{View selection} has been reduced based on Lemma~\ref{optimality properties}, the complexity of Algorithm 1 is still unacceptable when $K$ is large. In this part, we propose a low-complexity algorithm to obtain a suboptimal solution of Problem~\ref{View selection and resource allocation}.

First, by Lemma~\ref{optimality properties}, we can replace $x_v$ in Problem~\ref{View selection and resource allocation} with $\max_{k\in\mathcal{K}}y_{k,v}$ for all $v \in \overline{\mathcal{V}}$, without loss of optimality. Recalling that $\mathbf{p}$ can be determined by $\mathbf{e}$ and $\mathbf{t}$, we can use $\mathbf{e}$ instead of $\mathbf{p}$. The discrete constraints in (\ref{binary constraint y}) can be equivalently transformed to\textcolor{black}{:}
\begin{align}
&y_{k,v}\in [0,1],\quad k\in\mathcal{K},~v \in \overline{\mathcal{V}},\label{relaxed y}\\
&y_{k,v}(1-y_{k,v})\leq 0, \quad k\in\mathcal{K},~v\in \overline{\mathcal{V}}.\label{transformed binary <y}
\end{align}
\textcolor{black}{The reasons are given below. It is clear that (\ref{binary constraint y}) implies (\ref{relaxed y}) and (\ref{transformed binary <y}). By (\ref{relaxed y}), we have $y_{k,v}(1-y_{k,v})\geq 0,~k\in\mathcal{K},~v\in\overline{\mathcal{V}}$. Together with (\ref{transformed binary <y}), we have $y_{k,v}(1-y_{k,v})= 0,~k\in\mathcal{K},~v\in\overline{\mathcal{V}}$, implying (\ref{binary constraint y}). Therefore, (\ref{binary constraint y}) is equivalent to (\ref{relaxed y}) and (\ref{transformed binary <y}).} By noting that the constraints in (\ref{transformed binary <y}) are concave, we can disregard the constraints in (\ref{transformed binary <y}) and add to the objective function a penalty for violating them. Therefore, we can convert Problem~\ref{View selection and resource allocation} to the following problem.
\begin{problem}[Penalized DC Problem of Problem~\ref{View selection and resource allocation}]\label{Penalty relaxed problem}
\begin{align}
&\min_{\mathbf{y},\mathbf{t},\mathbf{e}}~ \textcolor{black}{\mathbb{E}_{\mathbf{H}}} \left[ \sum_{v \in \overline{\mathcal{V}}} e_{\mathbf{H},v}\right]+E_b\sum_{v\in \overline{\mathcal{V}}\setminus\mathcal{V}}\max_{k\in\mathcal{K}} y_{k,v}\nonumber\\
&\qquad \qquad \qquad \qquad \quad +\beta \sum_{k\in \mathcal{K}} (1-y_{k,r_k})E_{\text{u},k}+\rho P(\mathbf{y}) \notag\\
&~\text{s.t.}\quad(\ref{Right constraint}),(\ref{Left constraint}),(\ref{rest zero contraint}),(\ref{t>=0}),(\ref{time constraint}),(\ref{e>=0}),(\ref{transformed bandwidth constraint}),(\ref{relaxed y}), \notag
\end{align}
\end{problem}
where the penalty parameter $\rho>0$ and the penalty function $P(\mathbf{y})$ is given by
\begin{equation}
P(\mathbf{y})=\sum_{k\in \mathcal{K}}\sum_{v\in \overline{\mathcal{V}}} y_{k,v}(1-y_{k,v}). \label{penalty function}
\end{equation}

Note that the objective function of Problem~\ref{Penalty relaxed problem} can be viewed as a difference of two convex functions and the feasible set of Problem~\ref{Penalty relaxed problem} is convex. Thus, Problem~\ref{Penalty relaxed problem} can be viewed as a penalized DC problem of Problem~\ref{View selection and resource allocation}. By \cite{phan2012nonsmooth}, there exists $\rho_0>0$ such that for all $\rho>\rho_0$, Problem~\ref{Penalty relaxed problem} is equivalent to Problem~\ref{View selection and resource allocation}. Now, we solve Problem~\ref{Penalty relaxed problem} instead of Problem~\ref{View selection and resource allocation} by using the DC algorithm in \cite{lipp2016variations}. The main idea is to iteratively solve a sequence of convex approximations of Problem~\ref{Penalty relaxed problem}, each of which is obtained by linearizing the penalty function $P(\mathbf{y})$. Specifically, the convex approximation of Problem~\ref{Penalty relaxed problem} at the $i$-th iteration is given below.
\begin{problem}[Convex Approximation of Problem~\ref{Penalty relaxed problem} at $i$-th Iteration]\label{Convexified Penalty Problem}
\begin{align}
&(\mathbf{y}^{(i)},\mathbf{t}^{(i)},\mathbf{e}^{(i)}) \triangleq
\arg\min_{\mathbf{y},\mathbf{t},\mathbf{e}} \textcolor{black}{\mathbb{E}_{\mathbf{H}}} \left[ \sum_{v \in \overline{\mathcal{V}}} e_{\mathbf{H},v}\right]\nonumber\\
&+E_b\sum_{v\in \overline{\mathcal{V}}\setminus\mathcal{V}}\max_{k\in\mathcal{K}} y_{k,v}+\beta \sum_{k\in \mathcal{K}} (1-y_{k,r_k})E_{\text{u},k}+\rho\hat{P}\left(\mathbf{y};\mathbf{y}^{(i-1)}\right) \notag\\
&~\text{s.t.}\quad (\ref{Right constraint}),(\ref{Left constraint}),(\ref{rest zero contraint}),(\ref{t>=0}),(\ref{time constraint}),(\ref{transformed bandwidth constraint}),(\ref{relaxed y}), \notag
\end{align}
\end{problem}
where
\begin{align}
\hat{P}\left(\mathbf{y};\mathbf{y}^{(i-1)}\right)&\triangleq P\left(\mathbf{y}^{(i-1)}\right)+\nabla P\left(\mathbf{y}^{(i-1)}\right)^T \left(\mathbf{y}-\mathbf{y}^{(i-1)}\right) \notag \\
&=\sum_{k \in \mathcal{K}} \sum_{v\in \overline{\mathcal{V}}}  \left(1-2y_{k,v}^{(i-1)}\right)y_{k,v}+\left(y_{k,v}^{(i-1)}\right)^2. \notag
\end{align}
Here, $\mathbf{y}^{(i-1)}$ denotes an optimal solution of Problem~\ref{Convexified Penalty Problem} at the $(i-1)$-th iteration.

Problem~\ref{Convexified Penalty Problem} is a convex optimization problem and can be solved using standard convex optimization techniques. Similarly, to improve computation efficiency, we can adopt partial dual decomposition and parallel computation, as in Section~\ref{dual decomposition}. Due to space limitation, we omit the details.

It is known that the sequence \textcolor{black}{$\{(\mathbf{y}^{(i)},\mathbf{t}^{(i)},\mathbf{e}^{(i)})\}$} generated by the DC algorithm is convergent, and its limit point is a stationary point of Problem~\ref{Penalty relaxed problem}. \textcolor{black}{We can run the DC algorithm multiple times, each with a random initial feasible point of Problem~\ref{Penalty relaxed problem}. Then, we select the stationary point with the minimum average weighted sum energy among those with zero penalty, denoted by $\{(\mathbf{y}^{\dagger},\mathbf{t}^{\dagger},\mathbf{e}^{\dagger})\}$.} Due to the equivalence between Problems~\ref{View selection and resource allocation} and \ref{Penalty relaxed problem}, we know that for sufficiently large $\rho$, we can obtain a feasbile solution of Problem~\ref{View selection and resource allocation} based on $(\mathbf{y}^\dagger,\mathbf{t}^\dagger,\mathbf{e}^\dagger)$ as follows. Based on $\mathbf{y}^\dagger$, we obtain $\mathbf{x}^\dagger \triangleq (x_v^\dagger)_{v \in \overline{\mathcal{V}}}$ according to Lemma~\ref{optimality properties} (i). Based on $\mathbf{t}^\dagger$ and $\mathbf{e}^\dagger$, we then compute $\mathbf{p}^\dagger \triangleq (p_{\mathbf{h},v}^\dagger)_{v \in \overline{\mathcal{V}},\mathbf{h}\in\mathcal{H}^K}$ using $p^\dagger_{\mathbf{h},v} = e_{\mathbf{h},v}^\dagger/t_{\mathbf{h},v}^\dagger$. $(\mathbf{x}^\dagger,\mathbf{y}^\dagger,\mathbf{t}^\dagger,\mathbf{p}^\dagger)$ can serve as a suboptimal solution of Problem~\ref{View selection and resource allocation}. The details are summarized in Algorithm~2.\footnote{\textcolor{black}{Note that in Algorithm~2, $\mathbb{E}_{\mathbf{H}}\left[\sum_{v\in\overline{\mathcal{V}}} e_{\mathbf{H},v}^{(i)}\right]=\sum_{\mathbf{h}\in\mathcal{H}^K} q_{\mathbf{H}}(\mathbf{h})\sum_{v\in\overline{\mathcal{V}}} e_{\mathbf{H},v}^{(i)}$ can be computed as $q_{\mathbf{H}}(\mathbf{h}),\mathbf{h}\in\mathcal{H}^K$ are known.}}
\begin{algorithm}[t]
	\small
	\caption{Algorithm for Obtaining \textcolor{black}{a} Suboptimal Solution of Problem~\ref{View selection and resource allocation}}
	\textbf{Input} $c\geq 1$\\
	\textbf{Output} $(\mathbf{x}^\dagger,\mathbf{y}^\dagger,\mathbf{t}^\dagger,\mathbf{p}^\dagger)$
	\begin{algorithmic}[1] \label{DC Algorithm}
		\STATE $E=+\infty$.
		\WHILE{$c>0$}
		\STATE Find a random feasible point  of Problem~\ref{Penalty relaxed problem} as the initial point $(\mathbf{y}^{(0)},\mathbf{t}^{(0)},\mathbf{e}^{(0)})$, choose a sufficiently large $\rho$, and set $i=0$.
		\REPEAT
		\STATE Set $i=i+1$.
		\STATE Obtain $(\mathbf{y}^{(i)},\mathbf{t}^{(i)},\mathbf{e}^{(i)})$ of Problem~\ref{Convexified Penalty Problem} using standard convex optimization techniques or partial dual decomposition and parallel computation (similar to \textcolor{black}{Steps~3-7} in Algorithm~1).
		\UNTIL convergence criteria is met.
		 \IF{$P(\mathbf{y}^{(i)})=0$}
		\STATE Set $c=c-1$.
		\IF{$\textcolor{black}{\mathbb{E}_{\mathbf{H}}} \left[ \sum_{v \in \overline{\mathcal{V}}} e_{\mathbf{H},v}^{(i)}\right]+E_b\sum_{v\in \overline{\mathcal{V}}\setminus\mathcal{V}}\max_{k\in\mathcal{K}} y_{k,v}^{(i)}+\beta \sum_{k\in \mathcal{K}} (1-y_{k,r_k}^{(i)})E_{\text{u},k}<E$}
		\STATE Set $E=\textcolor{black}{\mathbb{E}_{\mathbf{H}}} \left[ \sum_{v \in \overline{\mathcal{V}}} e_{\mathbf{H},v}^{(i)}\right]+E_b\sum_{v\in \overline{\mathcal{V}}\setminus\mathcal{V}}\max_{k\in\mathcal{K}} y_{k,v}^{(i)}+\beta \sum_{k\in \mathcal{K}} (1-y_{k,r_k}^{(i)})E_{\text{u},k}$, $\mathbf{x}^\dagger=(x_v^\dagger)_{v\in\overline{\mathcal{V}}}$, $\mathbf{y}^\dagger = \mathbf{y}^{(i)}$, $\mathbf{t}^\dagger = \mathbf{t}^{(i)}$ and $\mathbf{p}^\dagger=(p^\dagger_{\mathbf{h},v})_{\mathbf{h}\in\mathcal{H}^K,v\in\overline{\mathcal{V}}}$, where $x_v^\dagger=\max y_{k,v}^{(i)}, v\in\overline{\mathcal{V}}$ and $p^\dagger_{\mathbf{h},v} = e_{\mathbf{h},v}^{(i)}/t_{\mathbf{h},v}^{(i)},\mathbf{h}\in\mathcal{H}^K,v\in\overline{\mathcal{V}}$.
		\ENDIF
		\ENDIF
		\ENDWHILE

	\end{algorithmic}
\end{algorithm}

\section{Total Utility Maximization} \label{Part II}
In this section, we consider the total utility maximization under the energy consumption constraints for the server and each user. We first formulate the optimization problem. Then, we develop a low-complexity algorithm to obtain a suboptimal solution using DC programming.
\subsection{Problem Formulation}
First, we impose the energy consumption constraints for the server and each user:
\begin{align}
&\textcolor{black}{\mathbb{E}_{\mathbf{H}}}\left[\sum_{v\in \overline{\mathcal{V}}} t_{\mathbf{H},v}p_{\mathbf{H},v}\right] + \sum_{v\in \overline{\mathcal{V}} \setminus \mathcal{V}}x_vE_b \leq \bar{E}_{b}, \label{bs energy constraint}\\
& (1-y_{k,r_k})E_{\text{u},k}\leq \bar{E}_{\text{u},k},\quad k \in \mathcal{K}, \label{user energy constraint}
\end{align}
where $\bar{E}_{b}$ and $\bar{E}_{\text{u},k}$ represent the energy consumption limits (for each time slot) at the server and user $k$, respectively. We would like to maximize the total utility by optimizing the view selection, transmission time and power allocation, and quality selection under the energy consumption constraints. Specifically, for given $\bar{E}_{b}$ and $\bar{E}_{\text{u},k},k\in\mathcal{K}$, we have the following optimization problem.
\begin{problem}[Total Utility Maximization]\label{original delta problem}
	\begin{align}
	&\max_{\mathbf{x},\mathbf{y},\mathbf{t},\mathbf{p},\boldsymbol{\Delta}} \quad U(\boldsymbol{\Delta}) \notag\\
	&~~\text{s.t.} \quad (\ref{Delta constraint}),(\ref{binary constraint x})-(\ref{bandwidth constraint}),(\ref{bs energy constraint}),(\ref{user energy constraint}).\notag
	\end{align}
\end{problem}
Let $(\mathbf{x}^\star,\mathbf{y}^\star,\mathbf{t}^\star,\mathbf{p}^\star,\boldsymbol{\Delta}^\star)$ denote an optimal solution of Problem~\ref{original delta problem} with slight abuse of notation, where $\boldsymbol{\Delta}^\star\triangleq (\Delta_k^\star)_{k\in\mathcal{K}}$.

Problem~\ref{original delta problem} is a challenging mixed discrete-continuous optimization problem with two types of variables, i.e., discrete view selection variables and quality selection variables $(\mathbf{x},\mathbf{y},\boldsymbol{\Delta})$ as well as continuous power and time allocation variables $(\mathbf{t},\mathbf{p})$. Problem~\ref{original delta problem} is even more \textcolor{black}{challenging} than Problem~\ref{View selection and resource allocation}, as it has extra discrete variables $\boldsymbol{\Delta}$ and the constraint functions of $\boldsymbol{\Delta}$ in (\ref{Right constraint})-(\ref{rest zero contraint}) are not tractable.

\subsection{Suboptimal Solution}
In this part, we propose a low-complexity algorithm to obtain a suboptimal solution of Problem~\ref{original delta problem} using equivalent transformations and DC programming. First, we transform Problem~\ref{original delta problem} to an equivalent penalized DC Problem which can be solved using DC programming. Specifically, we relax the discrete constraints in (\ref{Delta constraint}) to:
\begin{equation}
1\leq\Delta_k\leq V-1,\quad k \in \mathcal{K}. \label{relaxed Delta constraint}
\end{equation}
Then, we transform the constraints in (\ref{Right constraint})-(\ref{rest zero contraint}) to the following constraints:
\begin{align}
&y_{k,r_k}+\sum_{v>r_k,v\in \overline{\mathcal{V}}} y_{k,v}=1,\quad k\in \mathcal{K}, \label{new right constraint 1}\\
&v-r_k-\Delta_{k} \leq c(1-y_{k,v}),\quad v\in \overline{\mathcal{V}},\ k \in \mathcal{K},  \label{new right constraint 2}\\
&y_{k,r_k}+\sum_{v<r_k,v\in \overline{\mathcal{V}}} y_{k,v}=1,\quad k\in \mathcal{K}, \label{new left constraint 1}\\
&r_k-v-\Delta_{k} \leq c(1-y_{k,v}),\quad v\in \overline{\mathcal{V}},\ k \in \mathcal{K}, \label{new left constraint 2}
\end{align}
where $c>V-2$ is a positive constant. Next, as for solving Problem~\ref{View selection and resource allocation}, we eliminate $\mathbf{x}$, use $\mathbf{e}$ instead of $\mathbf{p}$, convert the discrete constraints in (\ref{binary constraint y}) to the continuous constraints in (\ref{relaxed y}) and (\ref{transformed binary <y}), and disregard (\ref{transformed binary <y}) by adding to the objective function a penalty for violating (\ref{transformed binary <y}). Therefore, we can convert Problem~\ref{original delta problem} to the following problem.
\begin{problem}[Penalized DC Problem of Problem~\ref{original delta problem}]
	\label{delta minimization}
	\begin{align}
	&\max_{\mathbf{y},\mathbf{t},\mathbf{e},\boldsymbol{\Delta}} \quad  U(\boldsymbol{\Delta})\textcolor{black}{-\rho P(\mathbf{y})} \notag\\
	&~~\text{s.t.} \quad (\ref{binary constraint x}),(\ref{t>=0}),(\ref{time constraint}),(\ref{e>=0}),(\ref{transformed bandwidth constraint}),(\ref{relaxed y}),(\ref{user energy constraint})-(\ref{new left constraint 2}) \notag \\
	&\quad \quad \quad \textcolor{black}{\mathbb{E}_{\mathbf{H}}} \left[ \sum_{v \in \overline{\mathcal{V}}} e_{\mathbf{H},v}\right] + E_b\sum_{v\in \overline{\mathcal{V}}\setminus\mathcal{V}}\max_{k\in\mathcal{K}} y_{k,v} \leq \bar{E}_{b}, \label{transformed bs energy constraint}
	\end{align}
\end{problem}
where $P(\mathbf{y})$ is given by (\ref{penalty function}). Let $(\mathbf{y}^*,\mathbf{t}^*,\mathbf{e}^*,\boldsymbol{\Delta}^*)$ denote an optimal solution of Problem~\ref{delta minimization}.

The following result shows the equivalence between Problem~\ref{original delta problem} and Problem~\ref{delta minimization}.
\begin{theorem}[Relationship between Problem~\ref{original delta problem} and Problem~\ref{delta minimization}]\label{theorem 1}
	There exists $\rho_0>0$ such that for all $\rho>\rho_0$ and $c>V-2$, $\mathbf{y}^\star=\mathbf{y}^*$, $\mathbf{t}^\star=\mathbf{t}^*$, $\mathbf{p}^\star=\mathbf{p}^*$ and $\boldsymbol{\Delta}^\star=\boldsymbol{\Delta}^*$, where $\mathbf{p}^*= (p^*_{\mathbf{h},v})_{\mathbf{h}\in \mathcal{H}^K,v\in\overline{\mathcal{V}}}$ with $p^*_{\mathbf{h},v}=e^*_{\mathbf{h},v}/t^*_{\mathbf{h},v},\mathbf{h}\in \mathcal{H}^K,v\in\overline{\mathcal{V}}$.
	\begin{IEEEproof}
		Please refer to Appendix C.
	\end{IEEEproof}
\end{theorem}

We can obtain a stationary point of Problem~\ref{delta minimization}, denoted by $(\mathbf{y}^\dagger,\mathbf{t}^\dagger,\mathbf{p}^\dagger,\boldsymbol{\Delta}^\dagger)$ with $P(\mathbf{y}^\dagger)=0$, using DC programming.
By Theorem~\ref{theorem 1}, we know that $(\mathbf{y}^\dagger,\mathbf{t}^\dagger,\mathbf{p}^\dagger,\boldsymbol{\Delta}^\dagger)$ is a feasible solution of Problem~\ref{original delta problem}. Similarly, based on $\mathbf{y}^\dagger$, we can obtain $\mathbf{x}^\dagger \triangleq (x_v^\dagger)_{v \in \overline{\mathcal{V}}}$ \textcolor{black}{with $x_v^\dagger=\max_{k\in\mathcal{K}} y_{k,v},v\in\overline{\mathcal{V}}$}. Based on $\mathbf{t}^\dagger$ and $\mathbf{e}^\dagger$, we then compute $\mathbf{p}^\dagger \triangleq (p_{\mathbf{h},v}^\dagger)_{v \in \overline{\mathcal{V}},\mathbf{h}\in\mathcal{H}^K}$ where $p^\dagger_{\mathbf{h},v} = e_{\mathbf{h},v}^\dagger/t_{\mathbf{h},v}^\dagger,\mathbf{h}\in \mathcal{H}^K,v\in\overline{\mathcal{V}}$. $(\mathbf{x}^\dagger,\mathbf{y}^\dagger,\mathbf{p}^\dagger,\mathbf{t}^\dagger,\boldsymbol{\Delta}^\dagger)$ serves as a suboptimal solution of Problem~\ref{original delta problem}. The details are  summarized in Algorithm~\ref{DC algorithm 2}.

\begin{algorithm}[t]
	\small
	\caption{Algorithm for Obtaining \textcolor{black}{a} Suboptimal Solution of Problem~\ref{original delta problem}}\label{DC algorithm 2}
	\textbf{Input} $c\geq 1$\\
	\textbf{Output} $(\mathbf{x}^\dagger,\mathbf{y}^\dagger,\mathbf{t}^\dagger,\mathbf{p}^\dagger,\boldsymbol{\Delta}^\dagger)$
	\begin{algorithmic}[1]
		\STATE Set $U=0$.
		\WHILE{$c>0$}
		\STATE Find a random feasible point  of Problem~\ref{original delta problem} as the initial point $(\mathbf{y}^{(0)},\mathbf{t}^{(0)},\mathbf{e}^{(0)},\boldsymbol{\Delta}^{(0)})$, choose a sufficiently large $\rho$ and $c>V-2$, and set $i=0$.
		\STATE Obtain $(\mathbf{y}^{(i)},\mathbf{t}^{(i)},\mathbf{e}^{(i)},\boldsymbol{\Delta}^{(i)})$ based on DC programming (similar to \textcolor{black}{Steps~4-7} in Algorithm~\ref{DC Algorithm}).
		\IF{$P(\mathbf{y}^{(i)})=0$}
		\STATE Set $c=c-1$.
		\IF{$U(\boldsymbol{\Delta}^{(i)})>U$}
		\STATE Set $U=U(\boldsymbol{\Delta}^{(i)})$, $\mathbf{x}^\dagger=(x_v^\dagger)_{v\in\overline{\mathcal{V}}}$, $\mathbf{y}^\dagger = \mathbf{y}^{(i)}$, $\mathbf{t}^\dagger = \mathbf{t}^{(i)}$, $\mathbf{p}^\dagger=(p^\dagger_{\mathbf{h},v})_{\mathbf{h}\in\mathcal{H}^K,v\in\overline{\mathcal{V}}}$ and $\boldsymbol{\Delta}^\dagger=\boldsymbol{\Delta}^{(i)}$, where $x_v^\dagger=\max_{k\in\mathcal{K}} y_{k,v}^{(i)}, v\in\overline{\mathcal{V}}$ and $p^\dagger_{\mathbf{h},v} = e_{\mathbf{h},v}^{(i)}/t_{\mathbf{h},v}^{(i)},\mathbf{h}\in\mathcal{H}^K,v\in\overline{\mathcal{V}}$.
		\ENDIF
		\ENDIF
		\ENDWHILE
		
	\end{algorithmic}
\end{algorithm}
\section{Numerical Results}
In the simulation, we set $\beta=2$, $R=18.59$~Mbit/s,\footnote{We use MVV sequence \emph{Kendo} as the video source \cite{video} and use HEVC in FFmpeg to encode the video with quantization parameter 15, frame rate 30 frame/s and resolution 1024$\times$768.} $E_b=10^{-3}$~Joule, $E_{\text{u},k}=10^{-3}$ Joule, $k \in \mathcal{K}$~\cite{6195536}, $\mathcal{V}=\{1,2,3,4,5\}$, $B=10$~MHz,\footnote{We consider a multi-carrier TDMA with 10 channels, each with bandwidth $1$~MHz.} $T=100$~ms, $U_k(\Delta_k)= V-\Delta_k, k \in \mathcal{K}$ and $\blue{\sigma^2}=Bk_BT_0$, where $k_B=1.38\times10^{-23}$~Joule/Kelvin is the Boltzmann constant and $T_0=300$~Kelvin is the temperature. For ease of simulation, we consider two channel states, i.e., a good channel state and a bad channel state, and set $\mathcal{H}=\{0.5d,1.5d\}$, $\Pr[H_k = 0.5d]=0.5$ and $\Pr[H_k = 1.5d]=0.5$ for all $k\in\mathcal{K}$, where $d=10^{-6}$ reflects the path loss. In addition, we assume that the $K$ users randomly request views in an i.i.d. manner. Specifically, for all $k\in\mathcal{K}$, $r_k$ falls in two regions, i.e., $\mathcal{V}_1\triangleq \{2,2+\frac{1}{Q},\cdots,4\}$ and $\mathcal{V}_2\triangleq \{1,1+\frac{1}{Q},\cdots,1+\frac{Q-1}{Q}\} \cup \{4+\frac{1}{Q},\cdots,5\}$, according to Zipf distribution with Zipf exponent $\gamma$,\footnote{Note that Zipf distribution is widely used to model content popularity in Internet and wireless networks. In addition, the proposed solutions and their properties are valid for arbitrary distributions of view requests.} i.e., $\Pr[r_k\in \mathcal{V}_1]=\frac{1^{-\gamma}}{\sum_{v\in\{1,2\}} v^{-\gamma}}\triangleq P_1$ and $\Pr[r_k\in \mathcal{V}_2]=\frac{2^{-\gamma}}{\sum_{v\in\{1,2\}} v^{-\gamma}}\triangleq P_2$. Note that a smaller $\gamma$ indicates a longer tail. Furthermore, for all $k\in\mathcal{K}$, a view in $\mathcal{V}_1$ or $\mathcal{V}_2$ is requested according to the uniform distribution, i.e., $\Pr[r_k=v]=\frac{P_1}{2Q+1}, v\in\mathcal{V}_1$ and $\Pr[r_k=v]=\frac{P_2}{2Q}, v\in\mathcal{V}_2$. We randomly generate 100 view requests for all users, and evaluate the average performance over these realizations.


For comparison, we consider two commonly used view selection mechanisms, based on which we shall construct optimization-based baseline schemes to minimize the weighted sum energy consumption and maximize the total utility, respectively. \textcolor{black}{In one view slection mechanism, the view requested by each user is transmitted \cite{zhao2015qos}. In our setup, this requires} view synthesis at the server but does not consider view synthesis at each user, and hence is referred to as the synthesis-server mechanism here. More specifically, for all $k\in\mathcal{K}$ and $v\in\overline{\mathcal{V}}$, $y_{k,v}=1$ if $v=r_k$, and $y_{k,v}=0$ otherwise; for all $v\in\overline{\mathcal{V}}$, $x_v=\max_{k\in\mathcal{K}}  y_{k,v}$. \textcolor{black}{In the other view selection mechanism, no synthesized views will be transmitted. In our setup, this requires} view synthesis at each user but does not consider view synthesis at the server~\cite{zhang2018packetization}, and hence is referred to as the synthesis-user mechanism here. More specifically, for all $k\in\mathcal{K}$, $y_{k,v}=0$ if $v\notin \textcolor{black}{\mathcal{V}\cap (\{r_k\} \cup \overline{\mathcal{V}}^-_{k,r_k}\cup \overline{\mathcal{V}}^+_{k,r_k})}$. We use Matlab and CVX toolbox to implement the proposed solutions and baseline schemes which are all optimization-based designs.


\subsection{Weighted Sum Energy Minimization}
\begin{figure}[t]	
	\subfigure[Weighted sum energy versus $K$.]{ 
		\begin{minipage}{0.46\linewidth}
			\centering
			\includegraphics[width=0.78\linewidth]{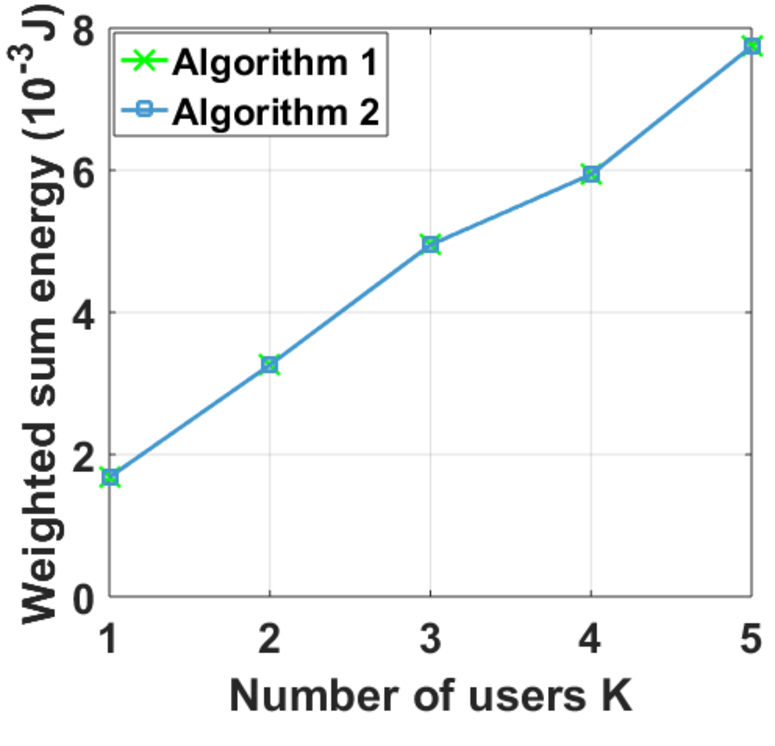}
			\vspace{0.5mm}
		\end{minipage}
	}
	\subfigure[Computation time versus $K$.]{
		\begin{minipage}{0.46\linewidth}
			\centering
			\includegraphics[width=\linewidth]{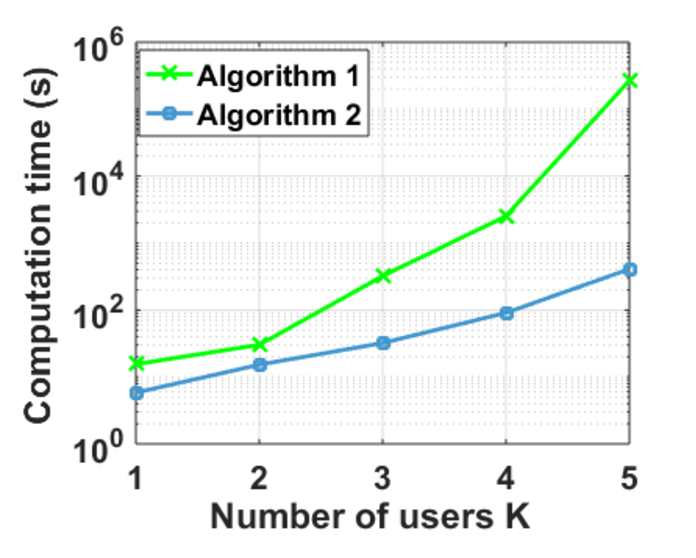}
		\end{minipage}
	}
	\caption{Comparison between the optimal solution and suboptimal solution at $\Delta_k=1$, $k\in\mathcal{K}$, $Q=5$ and $\gamma=1$.}\label{simulation_K}
\end{figure}

In this part, we compare the weighted sum energy consumptions of the proposed optimal and suboptimal solutions with those of two baseline schemes at $\Delta_k=1, k\in\mathcal{K}$. First, we compare the proposed optimal solution (obtained using Algorithm~1) with the proposed suboptimal solution of Problem~\ref{View selection and resource allocation} (obtained using Algorithm~\ref{DC Algorithm}) at small\footnote{Note that the computational complexity of Algorithm~1 is not acceptable when $K$ is large.} numbers of users. Fig.~\ref{simulation_K} illustrates the weighted sum energy consumption and computation time (reflecting computational complexity) versus the number of users, respectively. From Fig.~\ref{simulation_K}~(a), \textcolor{black}{we can see that the weighted sum energy consumption of the suboptimal solution is the same as that of the optimal solution when $K$ is small (under our setup).} From Fig.~\ref{simulation_K}~(b), we can see that the computation time of the suboptimal solution grows with the number of users at a much smaller rate than that of the optimal solution. This numerical example demonstrates the applicability and efficiency of the suboptimal solution.

Next, we compare the proposed suboptimal solution of Problem~\ref{View selection and resource allocation} with two baseline schemes, i.e., the synthesis-server-energy scheme and the synthesis-user-energy scheme. The two schemes adopt the synthesis-server mechanism and the synthesis-user mechanism, respectively. In addition, the synthesis-server-energy scheme adopts the optimal power and time allocation obtained by solving Problem~\ref{sub-problem for resource allocation} with $(\mathbf{x},\mathbf{y})$ chosen according to the synthesis-server mechanism using \textcolor{black}{Steps~3-13} of Algorithm~1; the synthesis-user-energy scheme adopts the power and time allocation and view selection obtained by solving Problem~\ref{View selection and resource allocation} with extra constraints on $\mathbf{y}$ which are imposed according to the synthesis-user mechanism using Algorithm~\ref{DC Algorithm}. Note that leveraging on our proposed transmission mechanism, both baseline schemes can utilize natural multicast opportunities; the synthesis-server-energy scheme does not create view synthesis-enabled multicast opportunities, but it can guarantee to transmit no more than $K$ views; the synthesis-user-energy scheme can create view synthesis-enabled multicast opportunities, but it may cause extra transmissions (i.e., may transmit more than $K$ views).

\begin{figure*}[t]
	\subfigure[Weighted sum energy versus $K$ at $\Delta_k=1$, $k\in\mathcal{K}$, $Q=5$ and $\gamma=1$.]{
		\begin{minipage}{0.3\linewidth}
			\centering
			\includegraphics[width=0.9\linewidth]{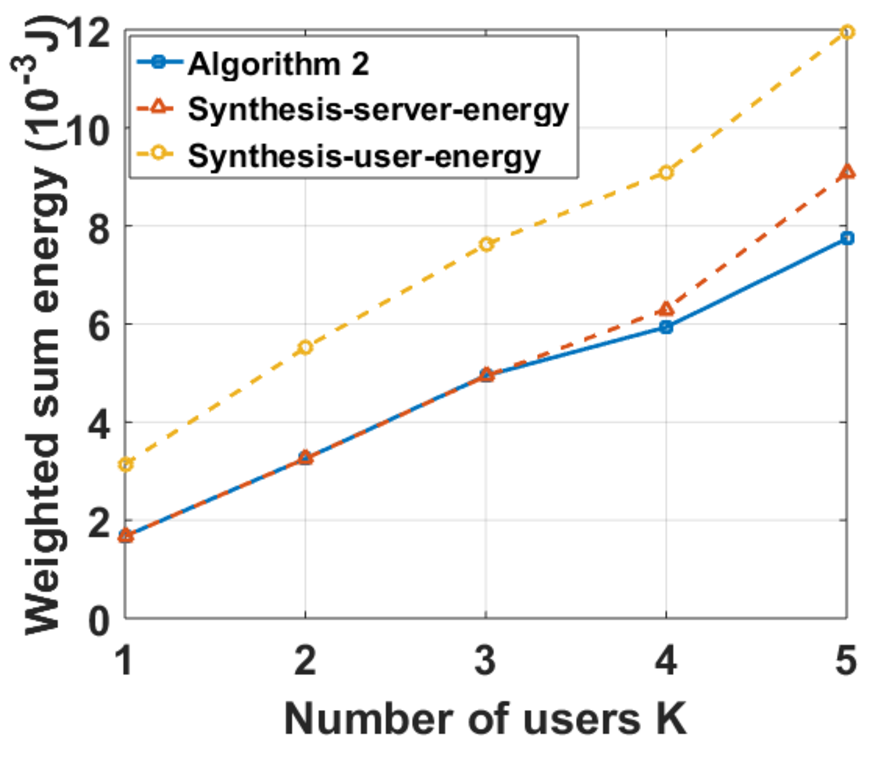}
		\end{minipage}
	}
	\subfigure[Weighted sum energy versus $\gamma$ at $\Delta_k=1$, $k\in\mathcal{K}$, $Q=5$ and $K=5$.]{
		\begin{minipage}{0.3\linewidth}
			\centering
			\includegraphics[width=0.9\linewidth]{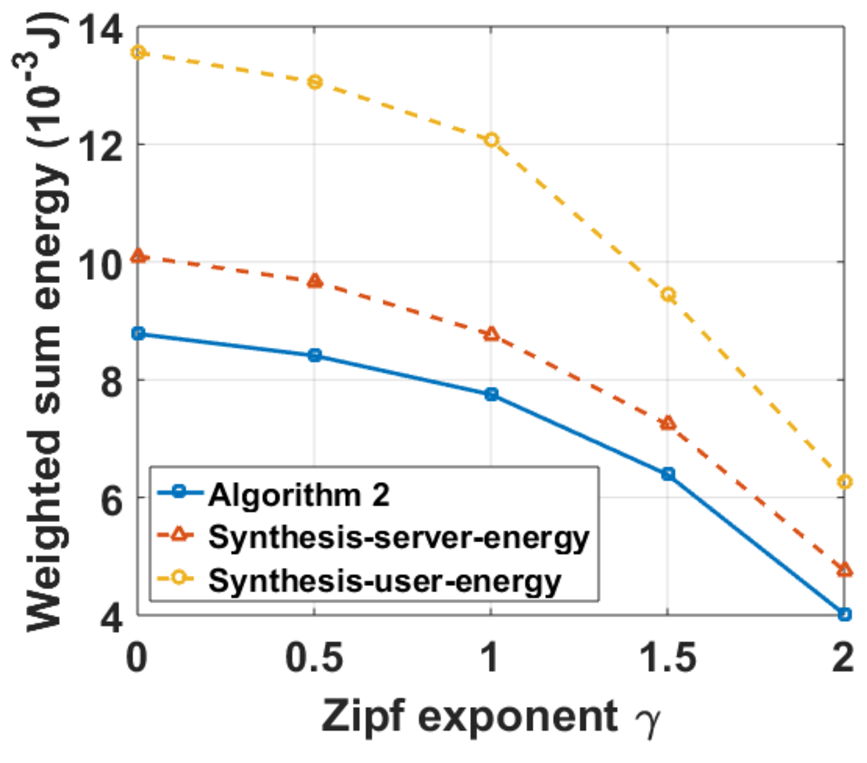}
 	\end{minipage}}
	\subfigure[Weighted sum energy versus $1/Q$ at $\Delta_k=1$, $k\in\mathcal{K}$, $\gamma=1$ and $K=5$.]{
		\begin{minipage}{0.3\linewidth}
			\centering
			\includegraphics[width=0.9\linewidth]{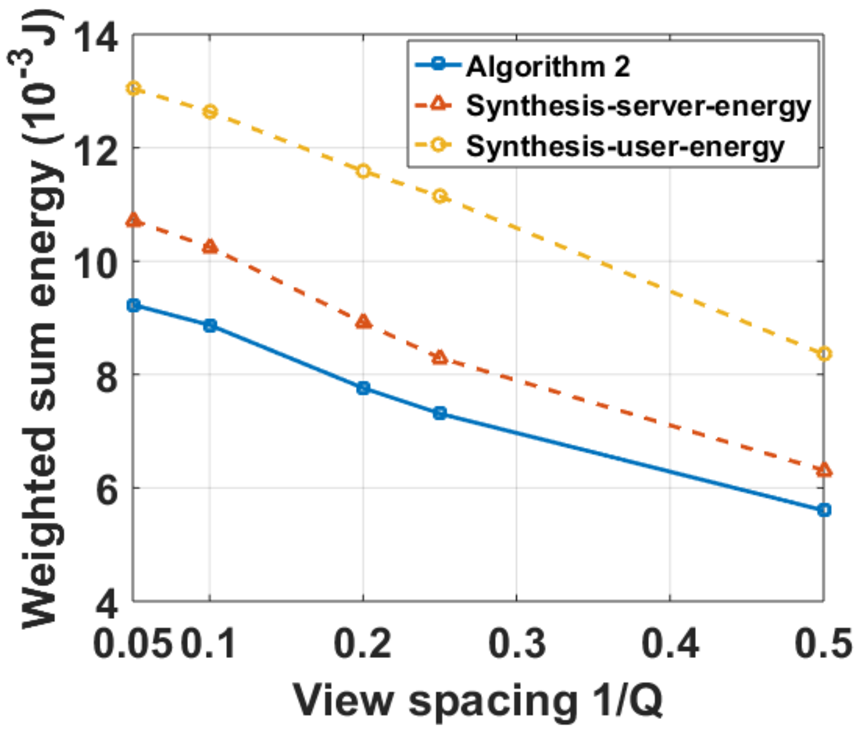}
	\end{minipage}}
	\caption{Comparison between the suboptimal solution and two baseline schemes.}
	\label{part I users}
\end{figure*}

Fig.~\ref{part I users} illustrates the weighted sum energy consumptions versus the number of users $K$, Zipf exponent $\gamma$ and view spacing $1/Q$. From Fig.~\ref{part I users}~(a), we can see that the weighted sum energy consumption of each scheme increases with $K$, as the traffic load increases with $K$. From Fig.~\ref{part I users}~(b) and Fig.~\ref{part I users}~(c), we can see that the weighted sum energy consumption of each scheme decreases with $\gamma$ and with $1/Q$. \textcolor{black}{This is because a larger $\gamma$ or a lager $1/Q$ indicates that view requests from the users are more concentrated, leading to more natural multicast opportunities}. From Fig.~\ref{part I users}, we can see that the synthesis-server-energy scheme outperforms the synthesis-user-energy scheme in most cases, demonstrating that creating view synthesis-enabled multicast opportunities in a naive manner usually causes extra transmissions and yields a higher weighted sum energy consumption; the proposed suboptimal solution outperforms the two baseline schemes, revealing the importance of the optimization of view synthesis-enabled multicast opportunities in reducing energy consumption. Note that the gains of the proposed suboptimal solution over the baseline schemes are large at large $K$, small $\gamma$ or small $1/Q$, as more view synthesis-enabled multicast opportunities can be created in these regions.

\subsection{Total Utility Maximization}
In this part, we compare the total utilities of the proposed suboptimal solution of Problem~\ref{original delta problem} (obtained using Algorithm~\ref{DC algorithm 2}) and two baselines, i.e., the synthesis-server-utility scheme and the synthesis-user-utility scheme, at $\bar{E}_b=10\times 10^{-3}$~Joule and $\bar{E}_{\text{u},k}=10^{-3}$~Joule, $k\in\mathcal{K}$.\footnote{Note that for each scheme, the constraints in (\ref{user energy constraint}) are always satisfied under the choices for $E_{\text{u},k},k\in\mathcal{K}$ and $\bar{E}_{\text{u},k},k\in\mathcal{K}$.} The two baseline schemes adopt the synthesis-server mechanism and the synthesis-user mechanism, respectively. In addition, the synthesis-server-utility scheme chooses \textcolor{black}{$\Delta_k=\min \{\Delta_k|(\ref{Delta constraint}),(\ref{Right constraint}),(\ref{Left constraint}),(\ref{rest zero contraint})\}$, $k\in\mathcal{K}$} with $(\mathbf{x},\mathbf{y})$ in (\ref{Right constraint}), (\ref{Left constraint}) and (\ref{rest zero contraint}) chosen according to the synthesis-server mechanism, and achieves total utility $U(\boldsymbol{\Delta})$ if Problem~\ref{original delta problem} with its choice for $(\mathbf{x},\mathbf{y},\boldsymbol{\Delta})$ is feasible; the synthesis-user-utility scheme achieves the total utility that is obtained by solving Problem~\ref{original delta problem} with extra constraints on $\mathbf{y}$ which are set according to the synthesis-user mechanism using Algorithm~\ref{DC algorithm 2}. Note that the synthesis-server-utility scheme and the synthesis-user-utility scheme share the same properties on natural and view synthesis-enabled multicast opportunities as the synthesis-server-energy scheme and the synthesis-user-energy scheme, respectively. For a realization of $r_k,k\in\mathcal{K}$, if the problem for each scheme is infeasible, we set the total utility to be 0, for ease of comparison.

\begin{figure*}[t]
	\subfigure[Total utility versus $K$ at $\gamma=1$, $Q=5$ and $\bar{E}_b=10\times 10^{-3}$~Joule and $\bar{E}_{\text{u},k}=10^{-3}$~Joule, $k\in\mathcal{K}$.]{
		\begin{minipage}{0.3\linewidth}	
			\centering
			\includegraphics[width=0.9\linewidth]{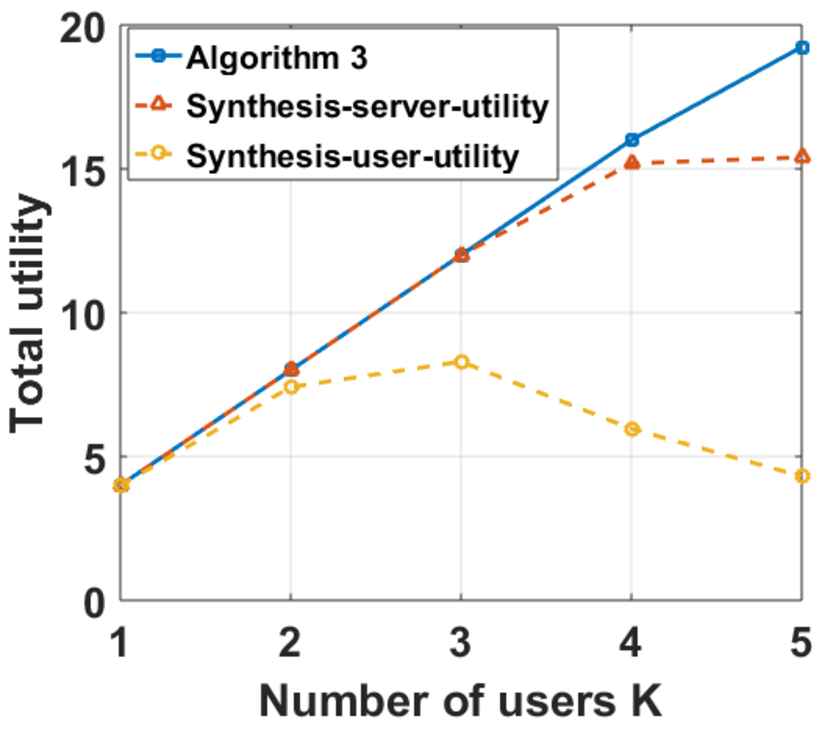}
		\end{minipage}
	}
	\subfigure[Total utility versus $\gamma$ at $K=5$, $Q=5$ and $\bar{E}_b=10\times 10^{-3}$~Joule and $\bar{E}_{\text{u},k}=10^{-3}$~Joule, $k\in\mathcal{K}$.]{
		\begin{minipage}{0.3\linewidth}
			\centering
			\includegraphics[width=0.87\linewidth]{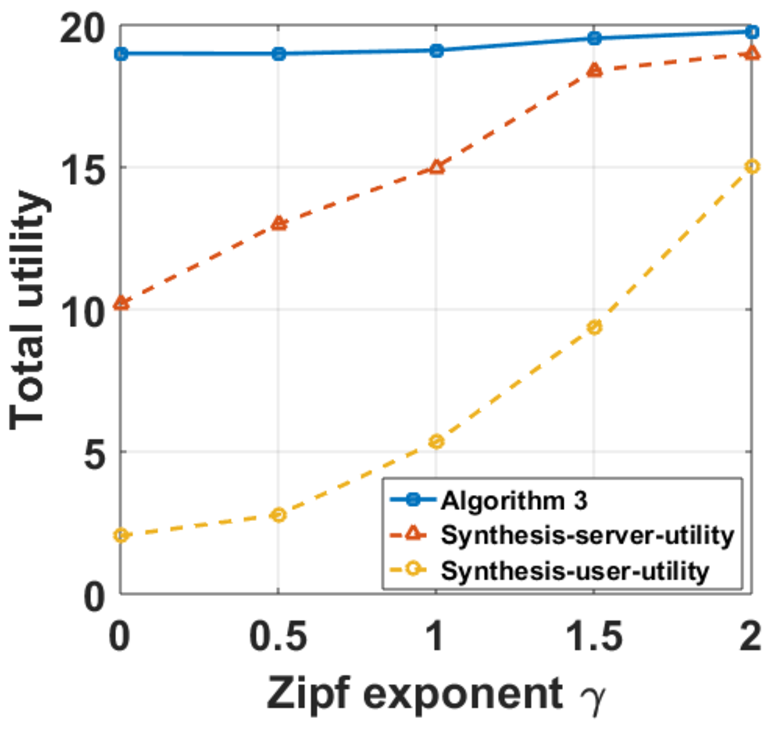}
		\end{minipage}
	}
	\subfigure[Total utility versus $1/Q$ at $\gamma=1$, $K=5$ and $\bar{E}_b=10\times 10^{-3}$~Joule and $\bar{E}_{\text{u},k}=10^{-3}$~Joule, $k\in\mathcal{K}$.]{
	\begin{minipage}{0.3\linewidth}
		\centering
		\includegraphics[width=0.88\linewidth,height=4.5cm]{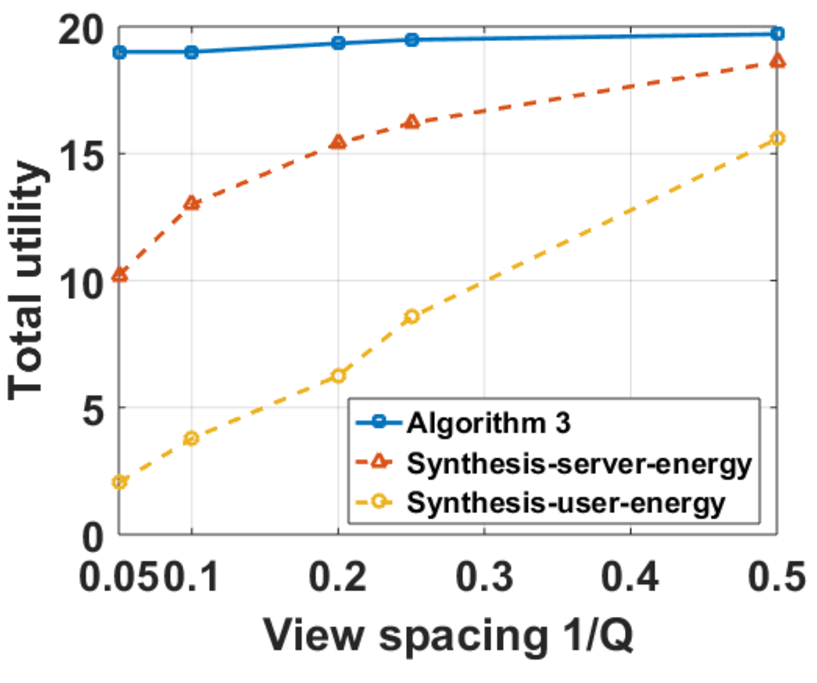}
	\end{minipage}
	}
	\caption{Comparison between the suboptimal solution and two baseline schemes.}
	\label{part II utility}
\end{figure*}

Fig.~\ref{part II utility} illustrates the total utility versus the number of users $K$, Zipf exponent $\gamma$ and view spacing $1/Q$. From Fig.~\ref{part II utility}~(a), we can see that the total utility of each scheme increases with $K$ when $K$ is small, as there is enough energy for serving more users; the total utilities of two baseline schemes \textcolor{black}{no longer increase} with $K$ when $K$ becomes large, as $\bar{E}_b$ and $\bar{E}_{\text{u},k}, k\in\mathcal{K}$ are not large enough and the optimization of the two baseline schemes are more likely to be infeasible under random user requests. From Fig.~\ref{part II utility}~(b) and Fig.~\ref{part II utility}~(c), we can see that the total utility of each scheme increases with $\gamma$ and with $1/Q$, as natural multicast opportunities increase with $\gamma$ and with $1/Q$. Finally, from Fig.~\ref{part II utility}, we see that the suboptimal solution outperforms the two baseline schemes, also revealing the importance of the optimization of view synthesis-enabled multicast opportunities in improving the total utility. Similarly, we can see that the gains of the proposed suboptimal solution over the baseline schemes are large at large $K$, small $\gamma$ or small $1/Q$, as more view synthesis-enabled multicast opportunities can be created in these regions.

\section{Conclusion}
In this paper, we considered optimal MVV transmission in a multiuser wireless network by exploiting both natural multicast opportunities and view synthesis-enabled multicast opportunities. First, we established a mathematical model to specify view synthesis at the server and each user and characterize its impact on multicast opportunities. To the best of our knowledge, this is the first mathematical model that enables the optimization of view synthesis-enabled multicast opportunities. \textcolor{black}{Then, we considered the minimization of the weighted sum energy consumption for view transmission and synthesis for given quality requirements of all users. We also considered the maximization of the total utility under the energy consumption constraints at the server and each user.} These two optimization problems are challenging mixed discrete-continuous optimization problems. We proposed an algorithm to obtain an optimal solution of the first problem with reduced computational complexity, by exploiting optimality properties. In addition, for each problem, we proposed a low-complexity algorithm to obtain a suboptimal solution, using DC programming. Finally, using numerical results, we showed the advantage of the proposed solutions, and demonstrated the importance of view synthesis-enabled multicast opportunities in MVV transmission.

\section*{Appendix A: Proof of Lemma~1}
First, we relax the coupling constraints in (\ref{transformed bandwidth constraint}) and obtain the following partial Lagrange function $L(\mathbf{y},\mathbf{t},\mathbf{e},\boldsymbol{\lambda})$, given by (\ref{Lagrangian function}), as shown at the top of the next page,
\begin{figure*}
\begin{align}
L(\mathbf{y},\mathbf{t},\mathbf{e},\boldsymbol{\lambda}) &\triangleq \sum_{\mathbf{h} \in \mathcal{H}^K} q_\mathbf{H}(\mathbf{h}) \sum_{v \in \overline{\mathcal{V}}} e_{\mathbf{h},v} - \sum_{k\in \mathcal{K}} \sum_{v \in\overline{\mathcal{V}}} \lambda_{k,v} \left(\frac{B}{T} \sum_{\mathbf{h}\in \mathcal{H}^K}  q_\mathbf{H}(\mathbf{h})  t_{\mathbf{h},v} \log_2\left(1+\frac{e_{\mathbf{h},v} h_k}{t_{\mathbf{h},v}\blue{\sigma^2}}\right) -y_{k,v}R\right) \notag\\
&= \sum_{\mathbf{h} \in \mathcal{H}^K} q_\mathbf{H}(\mathbf{h})  L_\mathbf{h}(\mathbf{y},\mathbf{t}_{\mathbf{h}},\mathbf{e}_{\mathbf{h}},\boldsymbol{\lambda}), \label{Lagrangian function}
\end{align}
\normalsize \hrulefill
\end{figure*}
where $\lambda_{k,v}, k\in \mathcal{K}, v\in\overline{\mathcal{V}}$ denote the Lagrange multipliers with respect to the constraints in (\ref{transformed bandwidth constraint}) and
$L_\mathbf{h}(\mathbf{y},\mathbf{t}_{\mathbf{h}},\mathbf{e}_{\mathbf{h}},\boldsymbol{\lambda}) \triangleq \sum_{v \in \overline{\mathcal{V}}} e_{\mathbf{h},v}- \sum_{k\in \mathcal{K}} \sum_{v \in \overline{\mathcal{V}}} \lambda_{k,v} \left(  \frac{B}{T} t_{\mathbf{h},v} \log_2\left(1+\frac{e_{\mathbf{h},v} h_k}{t_{\mathbf{h},v}\blue{\sigma^2}}\right) -y_{k,v}R \right).$ Next, for any given $\mathbf{y}\in\mathbf{Y}$, we obtain the corresponding partial dual function of Problem~\ref{sub-problem for resource allocation}:
\begin{align}
D(\mathbf{y},\boldsymbol{\lambda})\triangleq &\min_{\mathbf{t},\mathbf{e}} \quad  L(\mathbf{y},\mathbf{t},\mathbf{e},\boldsymbol{\lambda}) \notag\\
&~\text{s.t.} \quad (\ref{t>=0}),(\ref{time constraint}),(\ref{e>=0}),\notag
\end{align}
where $L(\mathbf{y},\mathbf{t},\mathbf{e},\boldsymbol{\lambda})$ is given by (\ref{Lagrangian function}). As the objective function and constraints are separable, this problem can be equivalently decomposed into Problem~\ref{decompostion}, one for each $\mathbf{h}\in\mathcal{H}^K$. As the duality gap for Problem~\ref{sub-problem for resource allocation} is zero, we can show Lemma~\ref{dual property}.
\section*{Appendix B: Proof of Lemma~2}
\subsection{Proof of Statement (i)}
We prove Statement~(i) by contradiction. Suppose there exists $v_0\in\overline{\mathcal{V}}$ such that $x_{v_0}^\star\neq \max_{k\in \mathcal{K}} y_{k,v_0}^\star$. By (\ref{x>y}), this implies $x_{v_0}^\star> \max_{k\in \mathcal{K}} y_{k,v_0}^\star$. By (\ref{binary constraint x}) and (\ref{binary constraint y}), we know $x_{v_0}^\star=1$ and $y_{k,v_0}^\star=0$, $k\in \mathcal{K}$. Construct $\mathbf{x}^\dagger\triangleq (x_v^\dagger)_{v\in\overline{\mathcal{V}}}$ with $x_{v_0}^\dagger=0$ and $x_v^\dagger=x_v^\star,v\neq v_0, v\in\overline{\mathcal{V}}$. It is clear that $\mathbf{x}^\dagger$ and $\mathbf{y}^\star$ satisfy (\ref{binary constraint x}) and (\ref{x>y}). In addition, the objective function of Problem~\ref{View selection and resource allocation}, i.e., $\textcolor{black}{\mathbb{E}_{\mathbf{H}}}\left[E(\mathbf{x},\mathbf{y},\mathbf{t}_\mathbf{H},\mathbf{p}_\mathbf{H})\right]$ increases with $x_v$, $v\in\overline{\mathcal{V}}$, and the constraints in (\ref{Right constraint}), (\ref{Left constraint}), (\ref{rest zero contraint}), (\ref{t>=0}), (\ref{time constraint}), (\ref{p>=0}), (\ref{bandwidth constraint}) do not rely on $\mathbf{x}$. Therefore, $(\mathbf{x}^\dagger,\mathbf{y}^\star,\mathbf{t}^\star,\mathbf{p}^\star)$ is a feasible solution with a smaller objective value than the optimal solution $(\mathbf{x}^\star,\mathbf{y}^\star,\mathbf{t}^\star,\mathbf{p}^\star)$. Therefore, by contradiction, we can prove Statement~(i).
\subsection{Proof of Statement (ii)}

We prove Statement~(ii) by contradiction. Suppose that there exist $k_0 \in \mathcal{K}$ and $v_0 \in \overline{\mathcal{V}} \setminus \left(\cup_{k\in\mathcal{K}} L_k\right)$ such that $ y^\star_{k_0,v_0} = 1$. \textcolor{black}{In the following, we consider three cases, i.e., $v_0 \notin \overline{\mathcal{V}}_{r_{k_0}}^- \cup \overline{\mathcal{V}}_{r_{k_0}}^+$, $v_0 \in \overline{\mathcal{V}}_{r_{k_0}}^+$ and $v_0 \in \overline{\mathcal{V}}_{r_{k_0}}^-$. In each case, we construct a feasible solution which achieves a smaller objective value under the assumption. Thus, by contradiction, we can show Statement~(ii).} 
\subsubsection{\textcolor{black}{$v_0 \notin \overline{\mathcal{V}}_{r_{k_0}}^- \cup \overline{\mathcal{V}}_{r_{k_0}}^+$}}
\textcolor{black}{First, we} construct $\mathbf{y}^\dagger\triangleq (y^\dagger_{k,v})_{k\in\mathcal{K},v\in\overline{\mathcal{V}}}$ with $y_{k_0,v_0}^\dagger=0$ and $y_{k,v}^\dagger = y_{k,v}^\star$, $(k,v)\neq (k_0,v_0),(k,v)\in \mathcal{K} \times \overline{\mathcal{V}}$, and $\mathbf{x}^\dagger\triangleq (x_v^\dagger)_{v\in\overline{\mathcal{V}}}$ with $x_{v_0}^\dagger=0$ and $x_{v}^\dagger=x_v^\star, v\neq v_0, v\in\overline{\mathcal{V}}$. It is easy to show that  $(\mathbf{x}^\dagger,\mathbf{y}^\dagger,\mathbf{t}^\star,\mathbf{p}^\star)$ is a feasible solution \textcolor{black}{with} a smaller objective value than $(\mathbf{x}^\star,\mathbf{y}^\star,\mathbf{t}^\star,\mathbf{p}^\star)$. 

\subsubsection{\textcolor{black}{$v_0 \in \overline{\mathcal{V}}_{r_{k_0}}^+$}}
First, define $\mathcal{K}_{v_0} \triangleq \{k\in \mathcal{K}~|y_{k,v_0} = 1\}$, $v_1 \triangleq \max_{k \in \mathcal{K}_{v_0}} f_{v_0}(r_k)$, and $\mathcal{M}_{v_0}\triangleq\{k\in \mathcal{K}_{v_0} | r_k=v_1\}$ where
\begin{equation*}
f_{v_0}(r_k)\triangleq \begin{cases}
r_k & r_k<v_0\\
r_k-\Delta & r_k>v_0
\end{cases}.
\end{equation*}
By (\ref{Right constraint}) and $y_{k,v_0}^\star=1 $, we know $y_{k,v_1}^\star=0$ for all $k \in \mathcal{K}_{v_0}$. Construct $\mathbf{y}^\dagger\triangleq (y_{k,v}^\dagger)_{k\in\mathcal{K},v\in\overline{\mathcal{V}}}$ with $y_{k,v_1}^\dagger=1,y_{k,v_0}^\dagger=0,y_{k,v}^\dagger=y_{k,v}^\star,v\in\overline{\mathcal{V}}\setminus\{v_0,v_1\}$, $k\in\mathcal{K}_{v_0}$ and $y_{k,v}^\dagger=y_{k,v}^\star,v\in\overline{\mathcal{V}},k\in\mathcal{K}\setminus \mathcal{K}_{v_0}$; construct $\mathbf{x}^\dagger\triangleq (x_v^\dagger)_{v\in\overline{\mathcal{V}}}$ with $x_{v}^\dagger=\max_{k\in\mathcal{K}} y_{k,v}^\dagger, v\in\overline{\mathcal{V}}$; construct $\mathbf{t}^\dagger \triangleq (t_{\mathbf{h},v}^\dagger)_{\mathbf{h}\in\mathcal{H}^K,v\in\overline{\mathcal{V}}}$ with $t_{\mathbf{h},v_0}^\dagger=0,t_{\mathbf{h},v_1}^\dagger=\max\{t_{\mathbf{h},v_0}^\star,t_{\mathbf{h},v_1}^\star\}$ and $t_{\mathbf{h},v}^\dagger=t_{\mathbf{h},v}^\star,v\in\overline{\mathcal{V}}\setminus\{v_0,v_1\}$, $\mathbf{h}\in\mathcal{H}^K$; construct $\mathbf{p}^\dagger\triangleq (p_{\mathbf{h},v})_{\mathbf{h}\in\mathcal{H}^K,v\in\overline{\mathcal{V}}}$ with $p_{\mathbf{h},v_0}^\dagger=0,p_{\mathbf{h},v_1}^\dagger=\max\{t_{\mathbf{h},v_0}^\star p_{\mathbf{h},v_0}^\star,t_{\mathbf{h},v_1}^\star p_{\mathbf{h},v_1}^\star\}/t_{\mathbf{h},v_1}^\dagger$ and $p_{\mathbf{h},v}^\dagger=p_{\mathbf{h},v}^\star,v\in\overline{\mathcal{V}}\setminus\{v_0,v_1\}$ for all $\mathbf{h}\in\mathcal{H}^K$. 

\textcolor{black}{Next, we show that $(\mathbf{x}^\dagger,\mathbf{y}^\dagger,\mathbf{t}^\dagger,\mathbf{p}^\dagger)$ is a feasible solution of Problem~\ref{View selection and resource allocation}.} It is clear that $(\mathbf{x}^\dagger,\mathbf{y}^\dagger,\mathbf{t}^\dagger,\mathbf{p}^\dagger)$ satisfies the constraints in (\ref{binary constraint x}), (\ref{binary constraint y}), (\ref{Right constraint}), (\ref{Left constraint}), (\ref{rest zero contraint}), (\ref{x>y}), (\ref{t>=0}) and (\ref{p>=0}). Then, we show that $(\mathbf{y}^\dagger,\mathbf{t}^\dagger,\mathbf{p}^\dagger)$ satisfies the constraints in (\ref{time constraint}) and (\ref{bandwidth constraint}). Since
$\sum_{v \in \overline{\mathcal{V}}}  t_{\mathbf{h},v}^\dagger \overset{(a)}{\leq} \sum_{v \in \overline{\mathcal{V}}} \nonumber
t_{\mathbf{h},v}^\star\leq T, ~\mathbf{h}\in\mathcal{H}^K,$
where (a) is due to  $t_{\mathbf{h},v_0}^\dagger=0,t_{\mathbf{h},v_1}^\dagger = \max \{t_{\mathbf{h},v_0}^\star,t_{\mathbf{h},v_1}^\star\},\mathbf{h}\in\mathcal{H}^K$ and $t_{\mathbf{h},v}^\dagger = t_{\mathbf{h},v}^\star,\mathbf{h}\in\mathcal{H}^K,v \in\overline{\mathcal{V}}\setminus\{v_0, v_1\} $, we can show that $\mathbf{t}^\dagger$ satisfies the constraints in (\ref{time constraint}). By the construction of $(\mathbf{x}^\dagger,\mathbf{y}^\dagger,\mathbf{t}^\dagger,\mathbf{p}^\dagger)$, \textcolor{black}{we know:}
\begin{align}
&\frac{B}{T}\textcolor{black}{\mathbb{E}_{\mathbf{H}}} \left[ t^\dagger_{\mathbf{H},v} \log_2\left(1+\frac{p^\dagger_{\mathbf{H},v} H_k}{\blue{\sigma^2}}\right) \right] \geq  y^\dagger_{k,v} R, \nonumber\\
&\qquad \qquad \qquad \qquad \qquad \qquad \quad k\in \mathcal{K},\  v\in \overline{\mathcal{V}}\setminus\{v_1\}\textcolor{black}{.} \label{prove_ii_1}
\end{align}
In addition, we have
\begin{align}
&\frac{B}{T} \textcolor{black}{\mathbb{E}_{\mathbf{H}}} \left[ t_{\mathbf{H},v_1}^\dagger \log_2\left(1+\frac{p_{\mathbf{H},v_1}^\dagger H_k}{\blue{\sigma^2}}\right) \right] \nonumber\\
&= \frac{B}{T} \textcolor{black}{\mathbb{E}_{\mathbf{H}}} \left[ t_{\mathbf{H},v_1}^\dagger \log_2\left(1+\frac{t_{\mathbf{H},v_1}^\dagger p_{\mathbf{H},v_1}^\dagger H_k}{t_{\mathbf{H},v_1}^\dagger \blue{\sigma^2}}\right) \right]  \nonumber\\
&\overset{(b)}{\geq} \frac{B}{T} \textcolor{black}{\mathbb{E}_{\mathbf{H}}} \left[  t_{\mathbf{H},v}^\star \log_2\left(1+\frac{t_{\mathbf{H},v}^\star p_{\mathbf{H},v}^\star H_k}{t_{\mathbf{H},v}^\star \blue{\sigma^2}}\right) \right], \nonumber\\
&\qquad \qquad \qquad \qquad \qquad \qquad \qquad\quad k\in\mathcal{K},~ v\in\{v_0,v_1\} \nonumber \\
\Rightarrow&\frac{B}{T} \textcolor{black}{\mathbb{E}_{\mathbf{H}}} \left[ t_{\mathbf{H},v_1}^\dagger \log_2\left(1+\frac{p_{\mathbf{H},v_1}^\dagger H_k}{\blue{\sigma^2}}\right) \right] \nonumber\\
&\geq\frac{B}{T} \max_{v\in\{v_0,v_1\}} \left\{ \textcolor{black}{\mathbb{E}_{\mathbf{H}}} \left[  t_{\mathbf{H},v}^\star \log_2\left(1+\frac{t_{\mathbf{H},v}^\star p_{\mathbf{H},v}^\star H_k}{t_{\mathbf{H},v}^\star \blue{\sigma^2}}\right) \right]\right\} \nonumber\\
&\overset{(c)}{\geq} R\max_{v\in\{v_0,v_1\}} y_{k,v}^\star  \overset{(d)}{=} y_{k,v_1}^\dagger R, \quad k\in\mathcal{K}, \label{prove_ii_3}
\end{align}
where (b) is due to that $x\log_2(1+y/x)$ is monotonically increasing with respect to $x$ and $y$, respectively, $t_{\mathbf{h},v_1}^\dagger p_{\mathbf{h},v_1}^\dagger \geq t_{\mathbf{h},v}^\star p_{\mathbf{h},v}^\star$, and $t_{\mathbf{h},v_1}^\dagger = t_{\mathbf{h},v}^\star$, $\mathbf{h}\in\mathcal{H}^K,v\in\{v_0,v_1\}$, (c) is due to the constraints in (\ref{bandwidth constraint}) for $(\mathbf{y}^\star,\mathbf{t}^\star,\mathbf{p}^\star)$, and (d) is due to $y_{k,v_1}^\dagger= y_{k,v_0}^\star=1$, $k\in\mathcal{K}_{v_0}$ and $y_{k,v_1}^\dagger=y_{k,v_1}^\star, y_{k,v_0}^\star=0$, $k\in\mathcal{K}\setminus \mathcal{K}_{v_0}$. By (\ref{prove_ii_1}) and (\ref{prove_ii_3}), we know that $(\mathbf{y}^\dagger,\mathbf{t}^\dagger,\mathbf{p}^\dagger)$ satisfies the constraints in (\ref{bandwidth constraint}). Thus, $(\mathbf{x}^\dagger,\mathbf{y}^\dagger,\mathbf{t}^\dagger,\mathbf{p}^\dagger)$ is a feasible solution of Problem~\ref{View selection and resource allocation}.

\textcolor{black}{Finally}, we prove $E^{\star}\geq E^{\dagger}\triangleq \textcolor{black}{\mathbb{E}_{\mathbf{H}}}\left[E(\mathbf{x}^{\dagger},\mathbf{y}^{\dagger},\mathbf{t}^{\dagger}_{\mathbf{H}},\mathbf{p}^{\dagger}_{\mathbf{H}})\right]$. By the construction of $(\mathbf{x}^\dagger,\mathbf{y}^\dagger,\mathbf{t}^\dagger,\mathbf{p}^\dagger)$, we have (\ref{pro2:ii}), which is shown at the top of the next page,
\begin{figure*}
\begin{align}
&E^{\star}-E^{\dagger}\nonumber\\
&=\textcolor{black}{\mathbb{E}_{\mathbf{H}}}\left[t_{\mathbf{H},v_0}^\star p_{\mathbf{H},v_0}^\star+t_{\mathbf{H},v_1}^\star p_{\mathbf{H},v_1}^\star -t_{\mathbf{H},v_0}^\dagger p_{\mathbf{H},v_0}^\dagger -t_{\mathbf{H},v_1}^\dagger p_{\mathbf{H},v_1}^\dagger \right]+(x_{v_0}^{\star}+x_{v_1}^{\star}-x_{v_0}^{\dagger}-x_{v_1}^{\dagger})E_b+\beta\sum_{k \in \mathcal{K}} \left(y^{\dagger}_{k,r_k}- y^{\star}_{k,r_k}\right)E_{\text{u},k} \notag\\
&=\textcolor{black}{\mathbb{E}_{\mathbf{H}}}\left[t_{\mathbf{H},v_0}^\star p_{\mathbf{H},v_0}^\star+t_{\mathbf{H},v_1}^\star p_{\mathbf{H},v_1}^\star -\max\{t_{\mathbf{H},v_0}^\star p_{\mathbf{H},v_0}^\star,t_{\mathbf{H},v_1}^\star p_{\mathbf{H},v_1}^\star\}- 0 \right]+(x_{v_0}^{\star}+x_{v_1}^{\star}-0-x_{v_1}^{\dagger})E_b+\beta \sum_{k\in \mathcal{M}_{v_0}} E_{\text{u},k} \nonumber\\
&\overset{(e)}{\geq} (x_{v_0}^\star-1)E_b + \beta \sum_{k\in\mathcal{M}_{v_0}}E_{\text{u},k}, \label{pro2:ii}
\end{align}
\normalsize \hrulefill
\end{figure*}
where (e) is due to $x+y-\max\{x,y\}\geq 0$ for all $x,y\geq0$ and $x_{v_1}^{\star}-x_{v_1}^{\dagger}\geq -1$. It remains to show $(x_{v_0}^\star-1)E_b + \beta \sum_{k\in\mathcal{M}_{v_0}}E_{\text{u},k}\geq 0$. We prove this by considering two cases.
\begin{itemize}
	\item \emph{Case 1:} $v_1\in \{r_k~ |~ k\in \mathcal{K}\}$. In this case, we have
	$
	(x_{v_0}^\star-1)E_b + \beta \sum_{k\in\mathcal{M}_{v_0}}E_{\text{u},k}\geq - E_b +\beta \sum_{k\in\mathcal{M}_{v_0}}E_{\text{u},k} \overset{(f)}{\geq} 0
	$
	where (f) is due to $ \mathcal{M}_{v_0}\neq \emptyset$ (as $v_1\in \{r_k~ |~ k\in \mathcal{K}\}$) and $-E_b+\beta E_{u,k}\geq 0$ for all $k \in \mathcal{K}$.
	\item \emph{Case 2:}
	$v_1\notin \{r_k~ |~ k\in \mathcal{K}\}$. First, by $v_1 = \max_{k \in \mathcal{K}_{v_0}} f_{v_0}(r_k)$, we know that there exists $k'\in\mathcal{K}_{v_0}$ such that $r_{k'} > v_0$. Thus we have $\overline{\mathcal{V}}_{r_{k_0}}^+ \cap \overline{\mathcal{V}}_{r_{k'}}^- \cap \mathcal{V}\subseteq L_{{k_0},{k'}}$ and $v_0 \in \overline{\mathcal{V}} \setminus \left(\cup_{k\in\mathcal{K}} L_k\right)$, implying that $v_0 \notin \overline{\mathcal{V}}_{r_{k_0}}^+ \cap \overline{\mathcal{V}}_{r_{k'}}^- \cap \mathcal{V}$. In addition, by noting that $y^\star_{k,v_0}=1,k\in\mathcal{K}_{v_0}$, we have \textcolor{black}{$x_{v_0}^\star=\max_{k\in\mathcal{K}} y_{k,v_0}=1$}. Thus, we have
	$
	(x_{v_0}^\star-1)E_b + \beta \sum_{k\in\mathcal{M}_{v_0}}E_{\text{u},k} \overset{(g)}{\geq} 0
	$\textcolor{black}{,} where (g) is due to $ \mathcal{M}_{v_0}= \emptyset$ (as $v_1\notin \{r_k~ |~ k\in \mathcal{K}\}$) and $x_{v_0}^\star = 1$.
\end{itemize}

Therefore, \textcolor{black}{$(\mathbf{x}^\dagger,\mathbf{y}^\dagger,\mathbf{t}^\star,\mathbf{p}^\star)$ is a feasible solution with a smaller objective value than $(\mathbf{x}^\star,\mathbf{y}^\star,\mathbf{t}^\star,\mathbf{p}^\star)$.}
\subsubsection{\textcolor{black}{$v_0 \in \overline{\mathcal{V}}_{r_{k_0}}^-$}} \textcolor{black}{The argument for $v_0 \in \overline{\mathcal{V}}_{r_{k_0}}^-$ is similar to \textcolor{black}{that for} $v_0 \in \overline{\mathcal{V}}_{r_{k_0}}^+$ and is omitted due to page limitation.}

\section*{Appendix C: Proof of Theorem~\ref{theorem 1}}
First, we show that Problem~\ref{original delta problem} and its continuous relaxation have the same optimal solution. By relaxing the discrete constraints in (\ref{Delta constraint}) into (\ref{relaxed Delta constraint}), we have:
\begin{problem}[Continuous Relaxation of Problem~\ref{original delta problem}]\label{relaxed delta problem}
	\begin{align}
	&\max_{\mathbf{x},\mathbf{y},\mathbf{t},\mathbf{p},\boldsymbol{\Delta}} \quad U(\boldsymbol{\Delta}) \notag\\
	&~~\text{s.t.} \quad (\ref{binary constraint x})-(\ref{bandwidth constraint}),(\ref{bs energy constraint}),(\ref{user energy constraint}),(\ref{relaxed Delta constraint}).\notag
	\end{align}
Let $(\mathbf{x}^\dagger,\mathbf{y}^\dagger,\mathbf{p}^\dagger,\mathbf{t}^\dagger,\boldsymbol{\Delta}^\dagger)$ denote an optimal solution of Problem~\ref{relaxed delta problem}.
\end{problem}

Note that the only difference between Problem~\ref{original delta problem} and Problem~\ref{relaxed delta problem} is that $\boldsymbol{\Delta}$ in Problem~\ref{original delta problem} satisfy (\ref{Delta constraint}) and $\boldsymbol{\Delta}$ in Problem~\ref{relaxed delta problem} satisfy (\ref{relaxed Delta constraint}). As the fact that $\boldsymbol{\Delta}$ satisfy (\ref{Delta constraint}) implies that $\boldsymbol{\Delta}$ satisfy (\ref{relaxed Delta constraint}), the optimal value of Problem~\ref{original delta problem} is no greater than the optimal value of Problem~\ref{relaxed delta problem}. Thus, to show that an optimal solution of Problem~\ref{relaxed delta problem} is also an optimal solution of Problem~\ref{original delta problem}, it remains to show that the optimal solution of Problem~\ref{relaxed delta problem} satisfies (\ref{Delta constraint}). We prove this by contradiction. Suppose $\boldsymbol{\Delta}^\dagger$ \textcolor{black}{does} not satisfy (\ref{Delta constraint}). Based on $\boldsymbol{\Delta}^\dagger$, we construct  $\underline{\boldsymbol{\Delta}}^\dagger\triangleq (\underline{\Delta}_k^\dagger)_{k\in\mathcal{K}}$, where $\underline{\Delta}_k^\dagger= \max \{x | x\leq \Delta_k^\dagger,(1)\}$.
It is clear that for all $k\in\mathcal{K}$,  $\epsilon_k\triangleq\Delta_k^\dagger-\underline{\Delta}_k^\dagger\in [0,1/Q)$. As  $U(\boldsymbol{\Delta})$ is a strictly decreasing function of $\boldsymbol{\Delta}$ and $\Delta_k^\dagger\geq \underline{\Delta}_k^\dagger, k \in \mathcal{K}$, we have  $U(\underline{\boldsymbol{\Delta}}^\dagger)\geq U(\boldsymbol{\Delta}^\dagger)$. \textcolor{black}{It remains to show that} $(\mathbf{x}^\dagger,\mathbf{y}^\dagger,\mathbf{p}^\dagger,\mathbf{t}^\dagger,\underline{\boldsymbol{\Delta}}^\dagger)$ is a feasible solution of Problem~\ref{relaxed delta problem}. Note that only the constraints in (\ref{Right constraint}), (\ref{Left constraint}) and (\ref{rest zero contraint}) involve $\boldsymbol{\Delta}$. Thus, it is sufficient to show for all $k\in\mathcal{K}$: (i) $\{x\in \overline{\mathcal{V}}: r_k<x\leq r_k+\Delta_k^\dagger\}=\{x\in \overline{\mathcal{V}}: r_k<x\leq r_k+\underline{\Delta}_k^\dagger\}$, (ii) $\{x\in \overline{\mathcal{V}}: r_k-\Delta_k^\dagger\leq x < r_k\}=\{x\in \overline{\mathcal{V}}: r_k-\underline{\Delta}_k^\dagger\leq x < r_k\}$, \textcolor{black}{and} (iii) $\{x\in \overline{\mathcal{V}}: x < r_k-\Delta_k^\dagger\}\cup \{x\in \overline{\mathcal{V}}:  r_k + \Delta_k^\dagger< x\}=\{x\in \overline{\mathcal{V}}: x < r_k-\underline{\Delta}_k^\dagger\}\cup \{x\in \overline{\mathcal{V}}:  r_k + \underline{\Delta}_k^\dagger< x\}$. We prove Case (i) as follows:
\begin{align}
&\{x\in \overline{\mathcal{V}}: r_k<x\leq r_k+\Delta_k^\dagger\} \nonumber\\
&= \{x\in \overline{\mathcal{V}}: r_k<x\leq r_k+\underline{\Delta}_k^\dagger+\epsilon_k\}\nonumber \\
&=\{x\in \overline{\mathcal{V}}: r_k<x\leq r_k+\underline{\Delta}_k^\dagger\}\nonumber\\
&\quad \quad\quad \quad\quad \quad\cup\{x\in \overline{\mathcal{V}}: r_k+\underline{\Delta}_k^\dagger<x\leq r_k+ \underline{\Delta}_k^\dagger+\epsilon_k\}\nonumber\\
&\overset{(a)}{=}\{x\in \overline{\mathcal{V}}: r_k<x\leq r_k+\underline{\Delta}_k^\dagger\} \cup \emptyset\nonumber\\
&=\{x\in \overline{\mathcal{V}}: r_k<x\leq r_k+\underline{\Delta}_k^\dagger\},
\end{align}
where (a) is due to that $r_k+\underline{\Delta}_k^\dagger\in\overline{\mathcal{V}}$, $\epsilon_k<1/Q$ and the view spacing is $1/Q$. Case (ii) and Case (iii) can be proved in a similar way to Case (i), and hence are omitted due to page limitation. Thus,  \textcolor{black}{we have shown that} $(\mathbf{x}^\dagger,\mathbf{y}^\dagger,\mathbf{p}^\dagger,\mathbf{t}^\dagger,\underline{\boldsymbol{\Delta}}^\dagger)$ is a feasible of Problem~\ref{relaxed delta problem} \textcolor{black}{with a larger objective value than $(\mathbf{x}^\dagger,\mathbf{y}^\dagger,\mathbf{p}^\dagger,\mathbf{t}^\dagger,\boldsymbol{\Delta}^\dagger)$, which contradicts the optimality of $(\mathbf{x}^\dagger,\mathbf{y}^\dagger,\mathbf{p}^\dagger,\mathbf{t}^\dagger,\boldsymbol{\Delta}^\dagger)$}. \textcolor{black}{Thus, we know} that Problem~\ref{original delta problem} and Problem~\ref{relaxed delta problem} have the same optimal solution.

Next, we show that Problem~\ref{relaxed delta problem} and the following problem have the same optimal solution.
\begin{problem}[Transformed Problem of Problem~\ref{original delta problem}]\label{Transformed Problem}
	\begin{align}
	&\max_{\mathbf{x},\mathbf{y},\mathbf{t},\mathbf{p},\boldsymbol{\Delta}} \quad U(\boldsymbol{\Delta}) \notag\\
	&~~\text{s.t.} \quad (\ref{binary constraint x}),(\ref{binary constraint y}),(\ref{x>y}),(\ref{t>=0}),(\ref{time constraint}),(\ref{p>=0}),(\ref{bandwidth constraint}),(\ref{bs energy constraint})-(\ref{new left constraint 2}).\notag
	\end{align}
\end{problem}
By comparing the constraints of Problem~\ref{relaxed delta problem} and \textcolor{black}{those of} Problem~\ref{Transformed Problem}, it is sufficient to show that the constraints in (\ref{binary constraint y}) and (\ref{new right constraint 1})-(\ref{new left constraint 2}) are equivalent to the constraints in (\ref{binary constraint y})-(\ref{rest zero contraint}), i.e., $\mathbf{Y}=\mathbf{Y}'$, where $\mathbf{Y}=\{\mathbf{y}:(\ref{binary constraint y}),(\ref{Right constraint}),(\ref{Left constraint}),(\ref{rest zero contraint})\}$ and $\mathbf{Y}'\triangleq\{\mathbf{y}:(\ref{binary constraint y}),(\ref{new right constraint 1}),(\ref{new right constraint 2}),(\ref{new left constraint 1}),(\ref{new left constraint 2})\}$. By (\ref{binary constraint y}) and (\ref{rest zero contraint}), we have $y_{k,v}=0, k\in\mathcal{K},v\in \{x\in \overline{\mathcal{V}}: x < r_k-\Delta_k\}\cup \{x\in \overline{\mathcal{V}}:  r_k + \Delta_k< x\}$. Thus, given that (\ref{binary constraint y}) and (\ref{rest zero contraint}) hold, (\ref{Right constraint}) and (\ref{Left constraint}) are equivalent to (\ref{new right constraint 1}) and (\ref{new left constraint 1}). Then, to show $\mathbf{Y}=\mathbf{Y}'$, it is sufficient to show $\{\mathbf{y}_k:(\ref{binary constraint y}),(\ref{rest zero contraint})\}=\{\mathbf{y}_k:(\ref{binary constraint y}),(\ref{new right constraint 2}),(\ref{new left constraint 2})\}$, for all $k\in\mathcal{K}$, where $\mathbf{y}_k \triangleq (y_{k,v})_{v\in\overline{\mathcal{V}}}$. We prove this by considering three cases for $k\in\mathcal{K}$: \textcolor{black}{(i) $v\in \overline{\mathcal{V}}_1 \triangleq \{x\in\overline{\mathcal{V}}:r_k-\Delta_k\leq v \leq r_k + \Delta_k\}$, (ii) $v\in\overline{\mathcal{V}}_2 \triangleq \{x\in\overline{\mathcal{V}}:x>r_k+\Delta_k\}$, and (iii) $v\in\overline{\mathcal{V}}_3 \triangleq \{x\in\overline{\mathcal{V}}:x<r_k-\Delta_k\}$.}
\begin{itemize}
\item \emph{Case (i):} In this case, as (\ref{rest zero contraint}) for $k$ is void, \textcolor{black}{$\{(y_{k,v})_{v\in\overline{\mathcal{V}}_1}:(\ref{binary constraint y}),(\ref{rest zero contraint})\}=\{(y_{k,v})_{v\in\overline{\mathcal{V}}_1}:(\ref{binary constraint y})\}$.} In addition, in this case, it is obvious that $v-r_k-\Delta_k \leq 0$ and $r_k-v-\Delta_k \leq 0$. \textcolor{black}{By (\ref{binary constraint y}) and $c>0$, we have $c(1-y_{k,v})\geq 0\geq v-r_k - \Delta_k$ and $c(1-y_{k,v})\geq 0\geq r_k-v - \Delta_k$.} Thus, in this case, (\ref{new right constraint 2}) and (\ref{new left constraint 2}) hold for $k$, implying \textcolor{black}{$\{(y_{k,v})_{v\in\overline{\mathcal{V}}_1}:(\ref{binary constraint y}),(\ref{new right constraint 2}),(\ref{new left constraint 2})\}=\{(y_{k,v})_{v\in\overline{\mathcal{V}}_1}:(\ref{binary constraint y})\}$.} Therefore, we can show \textcolor{black}{$\{(y_{k,v})_{v\in\overline{\mathcal{V}}_1}:(\ref{binary constraint y}),(\ref{rest zero contraint})\}=\{(y_{k,v})_{v\in\overline{\mathcal{V}}_1}:(\ref{binary constraint y}),(\ref{new right constraint 2}),(\ref{new left constraint 2})\}$.}
\item \emph{Case (ii):} In this case, \textcolor{black}{ $\{(y_{k,v})_{v\in\overline{\mathcal{V}}_2}:(\ref{binary constraint y}),(\ref{rest zero contraint})\}=\{(y_{k,v})_{v\in\overline{\mathcal{V}}_2}:y_{k,v}=0\}$.} \textcolor{black}{In this case,} \textcolor{black}{as $r_k-v-\Delta_k<r_k-v+\Delta_k<0$,} (\ref{new left constraint 2}) always holds for $k$. \textcolor{black}{Thus, it remains to show $\{(y_{k,v})_{v\in\overline{\mathcal{V}}_2}:y_{k,v}=0\}=\{(y_{k,v})_{v\in\overline{\mathcal{V}}_2}:(\ref{binary constraint y}),(\ref{new right constraint 2})\}$. First, we show that $\{(y_{k,v})_{v\in\overline{\mathcal{V}}_2}:(\ref{binary constraint y}),(\ref{new right constraint 2})\}$ implies $\{(y_{k,v})_{v\in\overline{\mathcal{V}}_2}:y_{k,v}=0\}$. By $r_k+\Delta_k \geq 2$, $c>V-2$ and $V\geq v$, we have $c>v-r_k-\Delta_k>0$, \textcolor{black}{which implies:}
\begin{equation}
0<\frac{v-r_k-\Delta_k}{c} <1. \label{proof statement (ii) 1}
\end{equation}
In addition, by (\ref{new right constraint 2}) and $c>V-2\geq 0$, we have:
\begin{equation}
y_{k,v} \leq 1 - \frac{v-r_k-\Delta_k}{c}. \label{proof statement (ii) 2}
\end{equation}
By (\ref{binary constraint y}), (\ref{proof statement (ii) 1}) and (\ref{proof statement (ii) 2}), we have $y_{k,v}=0,v\in\overline{\mathcal{V}}_2$. That is, $\{(y_{k,v})_{v\in\overline{\mathcal{V}}_2}:(\ref{binary constraint y}),(\ref{new right constraint 2})\}$ implies $\{(y_{k,v})_{v\in\overline{\mathcal{V}}_2}:y_{k,v}=0\}$. Next, it is obvious that  $\{(y_{k,v})_{v\in\overline{\mathcal{V}}_2}:y_{k,v}=0\}$ implies $\{(y_{k,v})_{v\in\overline{\mathcal{V}}_2}:(\ref{binary constraint y}),(\ref{new right constraint 2})\}$. Therefore, we can show $\{(y_{k,v})_{v\in\overline{\mathcal{V}}_2}:(\ref{binary constraint y}),(\ref{rest zero contraint})\}=\{(y_{k,v})_{v\in\overline{\mathcal{V}}_2}:(\ref{binary constraint y}),(\ref{new right constraint 2}),(\ref{new left constraint 2})\}$.}
\item \emph{Case (iii):} In this case, \textcolor{black}{$\{(y_{k,v})_{v\in\overline{\mathcal{V}}_3}:(\ref{binary constraint y}),(\ref{rest zero contraint})\}=\{(y_{k,v})_{v\in\overline{\mathcal{V}}_3}:y_{k,v}=0\}$.} \textcolor{black}{In this case,} \textcolor{black}{as $v-r_k-\Delta_k<v-r_k+\Delta_k<0$,} (\ref{new right constraint 2}) always holds for $k$. \textcolor{black}{Thus, it remains to show $\{(y_{k,v})_{v\in\overline{\mathcal{V}}_3}:y_{k,v}=0\}=\{(y_{k,v})_{v\in\overline{\mathcal{V}}_3}:(\ref{binary constraint y}),(\ref{new left constraint 2})\}$. First, we show that $\{(y_{k,v})_{v\in\overline{\mathcal{V}}_3}:(\ref{binary constraint y}),(\ref{new left constraint 2})\}$ implies $\{(y_{k,v})_{v\in\overline{\mathcal{V}}_3}:y_{k,v}=0\}$. By $v+\Delta_k \geq 2$, $c>V-2$ and $V\geq r_k$, we have $c>r_k-v-\Delta_k>0$, \textcolor{black}{which implies:}
\begin{equation}
0<\frac{r_k-v-\Delta_k}{c} <1. \label{proof statement (ii) 3}
\end{equation}
By (\ref{new left constraint 2}) and $c>V-2\geq 0$, we have:
\begin{equation}
y_{k,v} \leq 1 - \frac{r_k-v-\Delta_k}{c}. \label{proof statement (ii) 4}
\end{equation}
By (\ref{binary constraint y}), (\ref{proof statement (ii) 3}) and (\ref{proof statement (ii) 4}), we have $y_{k,v}=0,v\in\overline{\mathcal{V}}_3$. That is, $\{(y_{k,v})_{v\in\overline{\mathcal{V}}_3}:(\ref{binary constraint y}),(\ref{new left constraint 2})\}$ implies $\{(y_{k,v})_{v\in\overline{\mathcal{V}}_3}:y_{k,v}=0\}$. Next, it is obvious that  $\{(y_{k,v})_{v\in\overline{\mathcal{V}}_3}:y_{k,v}=0\}$ implies $\{(y_{k,v})_{v\in\overline{\mathcal{V}}_3}:(\ref{binary constraint y}),(\ref{new left constraint 2})\}$. Therefore, we can show $\{(y_{k,v})_{v\in\overline{\mathcal{V}}_3}:(\ref{binary constraint y}),(\ref{rest zero contraint})\}=\{(y_{k,v})_{v\in\overline{\mathcal{V}}_3}:(\ref{binary constraint y}),(\ref{new right constraint 2}),(\ref{new left constraint 2})\}$.}
\end{itemize}
Therefore, we can show that Problem~\ref{relaxed delta problem} and Problem~\ref{Transformed Problem} have the same optimal solution.

Finally, Problem~\ref{Transformed Problem} is a DC problem and Problem~\ref{delta minimization} can be viewed as its penalized DC problem. By \cite{phan2012nonsmooth}, we know that there exists $\rho_0 >0$ such that for all $\rho > \rho_0$, Problem~\ref{delta minimization} and Problem~\ref{Transformed Problem} have the same optimal solution.

Therefore, we complete the proof of Theorem~\ref{theorem 1}.

\end{document}